\documentclass[preprint]{elsarticle}
\usepackage[dvipsnames]{xcolor}

\usepackage{lineno}

\makeatletter
\def\ps@pprintTitle{%
 \let\@oddhead\@empty
 \let\@evenhead\
 \def\@oddfoot{}%
 \let\@evenfoot\@oddfoot}
\makeatother

\author[MMA]{Tomislav Mari\'{c}\corref{corr}}
\ead{maric@mma.tu-darmstadt.de}

\author[USTUTT]{Dennis Gl\"aser}
\ead{dennis.glaeser@iws.uni-stuttgart.de}

\author[SC]{Jan-Patrick Lehr}
\ead{jan-patrick.lehr@tu-darmstadt.de}

\author[MMA]{Ioannis Papagiannidis}
\ead{TODO}

\author[TD]{Benjamin Lambie}
\ead{lambie@tfi.tu-darmstadt.de}

\author[SC]{Christian Bischof}
\ead{christian.bischof@tu-darmstadt.de}

\author[MMA]{Dieter Bothe}
\ead{bothe@mma.tu-darmstadt.de}

\affiliation[MMA]{
 organization={Mathematical modeling and analysis, Mathematics department, TU Darmstadt},
 city={Darmstadt},
 postcode={64287}, 
 country={Germany}
}

\affiliation[USTUTT]{
 organization={Institute for Modelling Hydraulic and Environmental Systems, Department of Hydromechanics and Modelling of Hydrosystems, University of Stuttgart},
 city={Stuttgart},
 postcode={70569}, 
 country={Germany}
}

\affiliation[SC]{
 organization={Institute for Scientific Computing, Department of Computer Science, TU Darmstadt},
 city={Darmstadt},
 postcode={64289}, 
 country={Germany}
}

\affiliation[TD]{
 organization={Institute for Technical Thermodynamics, TU Darmstadt},
 city={Darmstadt},
 postcode={64287}, 
 country={Germany}
}

\cortext[corr]{Corresponding author}
\ead{maric@mma.tu-darmstadt.de}

\usepackage{amssymb}
\usepackage{textcomp}
\usepackage{mathtools}
\usepackage{subcaption}
\usepackage[section]{placeins}

\usepackage{lineno}

\usepackage[numbers]{natbib}
\usepackage{listings}

\usepackage{hyperref}
\usepackage{float}
\usepackage{algorithm}
\usepackage{graphicx}
\usepackage{algpseudocode}
\usepackage{amsmath}
\usepackage[size=footnotesize]{subcaption}
\usepackage[textsize=tiny,disable]{todonotes}
\graphicspath{{figures/}}
\usepackage{cleveref}
\usepackage{soul}
\usepackage{algorithmicx}
\usepackage{booktabs}
\usepackage{longtable}
\usepackage{stmaryrd}
\usepackage[inline]{enumitem}
\usepackage{xspace}
\usepackage{geometry}

\usepackage[utf8]{inputenc}

\newenvironment{enumi}{\begin{enumerate*}[label=(\arabic*)]}{\end{enumerate*}}

\newcommand{\cl}{\emph{C}}
\newcommand{\cppl}{\cl\texttt{++}}
\newcommand{\minwf}{minimum workflow}
\newcommand{\Minwf}{Minimum workflow}
\newcommand{\fuwf}{full workflow}
\newcommand{\Fuwf}{Full workflow}

\newcommand{\dg}[1]{\color{black}#1\color{black}\xspace}

\newcommand{\tm}[1]{\textcolor{black}{#1}}
\newcommand{\eg}{e.g.\@\xspace}

\title{A Research Software Engineering Workflow for Computational Science and Engineering}

\begin{document}

\begin{abstract}

University research groups in Computational Science and Engineering (CSE) generally lack dedicated funding and personnel for Research Software Engineering (RSE), which, combined with the pressure to maximize the number of scientific publications, shifts the focus away from sustainable research software development and reproducible results. The neglect of RSE in CSE at University research groups negatively impacts the scientific output: research data - including research software - related to a CSE publication cannot be found, reproduced, or re-used, different ideas are not combined easily into new ideas, and published methods must very often be re-implemented to be investigated further. This slows down CSE research significantly, resulting in considerable losses in time and, consequentially, public funding.

We propose a RSE workflow for Computational Science and Engineering (CSE) that addresses these challenges, that improves the quality of research output in CSE. Our workflow applies established software engineering practices adapted for CSE: software testing, result visualization, and periodical cross-linking of software with reports/publications and data, timed by milestones in the scientific publication process. The workflow introduces minimal work overhead, crucial for university research groups, and delivers modular and tested software linked to publications whose results can easily be reproduced. We define research software quality from a perspective of a pragmatic researcher: the ability to quickly find the publication, data, and software related to a published research idea, quickly reproduce results, understand or re-use a CSE method, and finally extend the method with new research ideas.

\end{abstract}


\maketitle

\thispagestyle{plain}
\pagestyle{plain}



\section{Introduction}
\label{sec:intro}

Research software \dg{engineering} - crucial in \textbf{Computational} Science and Engineering - continues to \dg{find little application in academia, potentially due to beliefs that the costs outweigh the benefits}~\citep{2020:heroux:psip}.
\dg{However,} efficient and long-term CSE research is impossible without sustainable research software development: without modularity and adequate test coverage, straightforward integration of different research ideas, and the ability to find and reproduce research results from publications, many CSE research projects stall.

Ensuring the sustainability of research software is often challenging even for a single researcher - often a Ph.D. student. Focusing only on \dg{an upcoming} scientific publication means focusing on the next set of computations, often wholly disregarding the impact of changes in the research software on the existing code base.
Adding new functionality while ensuring that complex existing CSE methods continue to work as they should is more complicated when multiple researchers work together on the same code base.
Sustainable research software, therefore, must allow the integration of contributions from different researchers and ensuring that previous contributions work as expected. 
The software industry has already established principles, workflows, and tools to cope with 
\tm{sustainable software development}.
These principles and workflows are widespread in larger scientific projects; however, they find little to no application in many smaller projects, particularly at universities.
In research, everything revolves around scientific publications. 
There is a need to \textbf{F}ind research software and data related to a publication and \textbf{A}ccess it. 
Once results are found and accessed,  there is a need to \textbf{I}nteract with them easily - to draw new conclusions, perform a more detailed analysis, \tm{apply the method outside of the original application-range}.
Finally, to extend or improve methods, being able to \dg{\textbf{R}euse and/or} \textbf{R}eproduce results from specific milestones is necessary.
The requirements mentioned above are the basis of the \href{https://www.go-fair.org/fair-principles/}{FAIR} principles \citep{FAIR}.
For reproducing results, the specific software configuration and its environment (dependencies), complete input data, and a result evaluation workflow are necessary.
Moreover, structured management of \emph{primary data} (\eg simulation results) and \emph{secondary data} (\eg diagrams and tables) is also necessary.




Access to the software and its configuration used in a publication is crucial for result reproduction. 
Without access to the CSE software, and its primary and secondary data, the only way to assess whether a numerical method is working is to re-implement it.
Re-implementation of published CSE results is extremely difficult - \tm{CSE methods are generally complex}, and not all details are available in a scientific publication.
\tm{The need to re-implement published work is slowing} down research in CSE and introducing equally significant financial losses in public research funding.



As for the challenge of developing sustainable research software, we argue that a careful integration of new features, involving automatic verification and validation tests must be done, and can be done in a relatively straightforward way in \tm{the university research environment}.
Without integration, research software is often developed in diverging directions, and integrating features quickly becomes intractable.
Computational sciences pose an additional challenge: large-scale tests that require high-performance computing are often necessary for verification and validation of the research software. 
This requires careful design of certain
parts of the workflow.


In this paper, we propose a workflow that addresses these challenges mostly based on standard software development practices applied to research software in CSE.
The workflow is lightweight and is focused on small research teams or individual researchers at universities, because university researchers seldom have dedicated personnel and resources that support the development of sustainable research software. 
In addition, we show at which points the workflow can be extended.






\begin{table*}[!hbt]
		\begin{tabular}{@{} *6l @{}}    \toprule
				\emph{Publication} & \emph{Branching model} & \emph{TDD} & \emph{Cross-linking} & \emph{CI}  & (Meta)data standardization \\\midrule
				 \citep{Jimenez2017}
					 & -  & -  & -  & - & 1  \\ 
				 \citep{Fehr2016} 
					 & -  & -  & -  & - & 2  \\ 
				 \citep{Wilson2014} 
					 & 1  & 2  & -  & - & -  \\ 
				 \citep{wilson_good_2017}
					 & -  & -  & 3 & 1 & 3  \\ 
				 \citep{stanisic_effective_2015}
					 & 1  & -  & - & 1 & -  \\ 
				 \citep{anzt_towards_2019}
					 & 1  & -  & - & 5 & -  \\ 
				 \citep{riesch_bertha_2020}
					 & 1  & -  & - & 1 & 4  \\ 
				 \citep{sampedro_continuous_2018}
					 & 1  & -  & -  & 4 & - \\
					 \bottomrule
				 \hline
		\end{tabular}
		\caption{A comparison of the proposed workflow with other workflows. Similarity is expressed subjectively using numbers 1 (not similar) to 5 (very similar), or '-' for an aspect that is not addressed.}
		\label{table:comparison}
\end{table*}

Increasing the quality of research software is gaining a lot of traction in the CSE community. 
Initiatives such as Better Scientific Software (BSSw) \cite{BSSw} and its German and British equivalents, the National Research Data Initiative (NFDI) \citep{NFDI} (with a community-driven knowledge-base \citep{NFDI4IngKBase}) and the Sustainable Software Initiative \citep{SSI}, support sustainable research software development by organizing workshops and providing best practices to the CSE community.
\tm{The "Turing Way" handbook \citep{TuringWay} is a community-driven excellent source of information on increasing the eproducibility of research results in data science.}

A discussion on best practices to increase the quality of scientific software can be found in \citep{Wilson2014}, which comprise quality aspects of the source code (\eg consistent naming and formatting), quality assurance measures (automated build and test pipelines), design principles (\eg "don't-repeat-yourself") and technical aspects such as the use of version control systems.

In \cite{stanisic_effective_2015}, the issue of reproducibility is addressed by combining a "git" version control workflow that includes research data, and Org-mode for the Emacs text editor research documentation. The workflow is minimalistic, but it is tied to the Emacs editor.

A distinction between replicability, reproducibility and reusability of research software is described in \cite{Fehr2016}. According to the authors, replicability of a computational experiment requires sufficient documentation on how to set up and run the program, while the automation and testing of this procedure is only recommended. Reproducibility of the results requires additional information on hardware and software dependencies, while reusability further demands modular software and a licensing model. The workflow presented in this work places automation at its core, with the aim of increasing the reproducibility. 

Making software publicly accessible as early as possible, making it easily found, applying a license compatible with its dependencies and ensuring transparent contribution and communication channels become relevant when research software attracts collaboration partners and starts growing into an open-source project \cite{Jimenez2017}. Moreover, we argue that early public access represents a psychological motivational factor for university research groups - Ph.D. students that are mainly responsible for new feature development, in our experience, place more focus on code quality if they are working on a publicly accessible code.

Continuous Integration (CI) for research software on HPC systems was proposed in \citep{sampedro_continuous_2018}, using a dedicated Jenkins test server in combination with a Singularity \citep{Singularity} registry  to ensure reproducibility of research results. This approach is very promising, especially regarding the use of Singularity images that encapsulate the computing environment of the research software. Since then, web-based version control services such as GitHub and GitLab offer Continuous Integration - online automatic testing systems, that are easy to set up and make the dedicated Jenkins test server unnecessary. 

Good enough practices in scientific computing according to \citep{wilson_good_2017} focus on data management, software, collaboration and project organization. Regarding data management, they recommend the archiving of
raw data while making it easily understandable.
Their software recommendations focus on code quality in terms of comments, function naming, etc., while their collaboration recommendations may nowadays be replaced by issue tracking available on git hosting sites. Good enough is a crucial attribute for small university research groups: a workflow must be simple enough to be quickly adopted by a small team of Ph.D. students and postdocs. 


Continuous Benchmarking (CB) is a useful method to make sure that the performance of a software does not deteriorate over time.
A CB workflow that includes version control, a modern build system, automatic testing framework, code reviews, benchmark tests on HPC systems, storage of performance data and their automatic visualization is presented in \citep{anzt_towards_2019}. They realized all the aforementioned steps by relying on available web-based technologies, apart from result visualization: the authors have developed their own web-based solution for this task.

For many research software projects, performance may be a secondary goal that comes after ensuring that physically meaningful results are being computed. To this end, Continuous Verification and Validation (VV) may be done by comparing test results against previous milestones, other research software, experiments or analytical solutions where available.

The importance of automatic visualization of results and their communication to team members is emphasized in \citep{Khuvis2019}. The authors use a custom workflow to manage continuous integration of research software using git hooks that trigger automatic tests and email as means of communication of test results.


Jupyter \citep{Jupyter} represents a powerful CSE environment for literate programming - interleaving documentation and executable code in Jupyter notebooks makes it possible to combine test documentation (problem description, method description) with data generation (simulations, experimental data processing) and visualization. In \citep{Beg2021}, Jupyter notebooks are combined with Binder \citep{Binder} - a web service for building custom computing environments that contain the software and all its dependencies. Focus is placed on using Jupyter Notebooks and literate programming to initialize and perform computational studies and visualize results. This workflow is relevant regarding reproducing results and interleaving computation and documentation.

A tabular summary of the comparison between our proposed workflow and those reported so far in the literature is given in \cref{table:comparison}.

Our minimalistic workflow applies all elements of the workflows mentioned above in a way that makes it easily adoptable by small university research groups in CSE and applicable also to legacy research software. 
We rely entirely on existing web services and open-source software for version control, automatic testing, computing environments, data management, and visualization. 
Another distinctive feature of our workflow is the focus on the scientific publication and the peer-review process as a source of milestones that mark the points in research that require the reproduction of results. In an ideal world, integration and testing are continuous, however, conditions for research software development at universities are far from ideal. Tying integration and testing to writing a scientific paper (submission, revisions, acceptance) significantly reduces the workload overhead of university researchers while still ensuring sustainable research software development. 

The paper is structured as follows: first, we outline an overview of the general workflow in \cref{sec:workflow}.
\dg{\Cref{sec:minimal-workflow,sec:full-workflow} outline a basic and full version of our workflow presenting their different aspects in detail, that is,
which challenge is addressed, which technology is applied, what are potential impediments, and how to overcome them}.
We \dg{describe a minimal working example in~\cref{sec:minimal-example}, while} two case studies are presented in \cref{sec:case-studies} that show the positive impact when following the suggested workflow.
Finally, we draw conclusions in \cref{sec:conclusions} and outline further activities required.

\section{A workflow for increasing the quality of scientific software}
\label{sec:workflow}

\todo[inline]{PASC review: consider reordering points to better introduce the reader to the workflow - a user story would help more.}

\todo[inline]{PASC review: figure 1 hard to interpret, see reviewer 3 response}

\todo[inline]{PASC review: unclear why sustainable software  and accessing and reproducing results need to be resolved.}

\todo[inline]{PASC: re-design of the workflow on HPC systems is not discussed in detail.}

\todo[inline]{PASC review: Integration alone with some testing is not sufficient for sustainable software: limited resources, university research groups, we are not saving the planet here.}

\todo[inline]{@Dennis: should we rename TDD to RQ-Driven-Development? Yes, not sure if I can do this before 06.08.}

\begin{figure}[tb]
    \centering
    \includegraphics[clip=true,trim={1.3cm 20.5cm 3cm 0.1cm},width=0.6\linewidth]{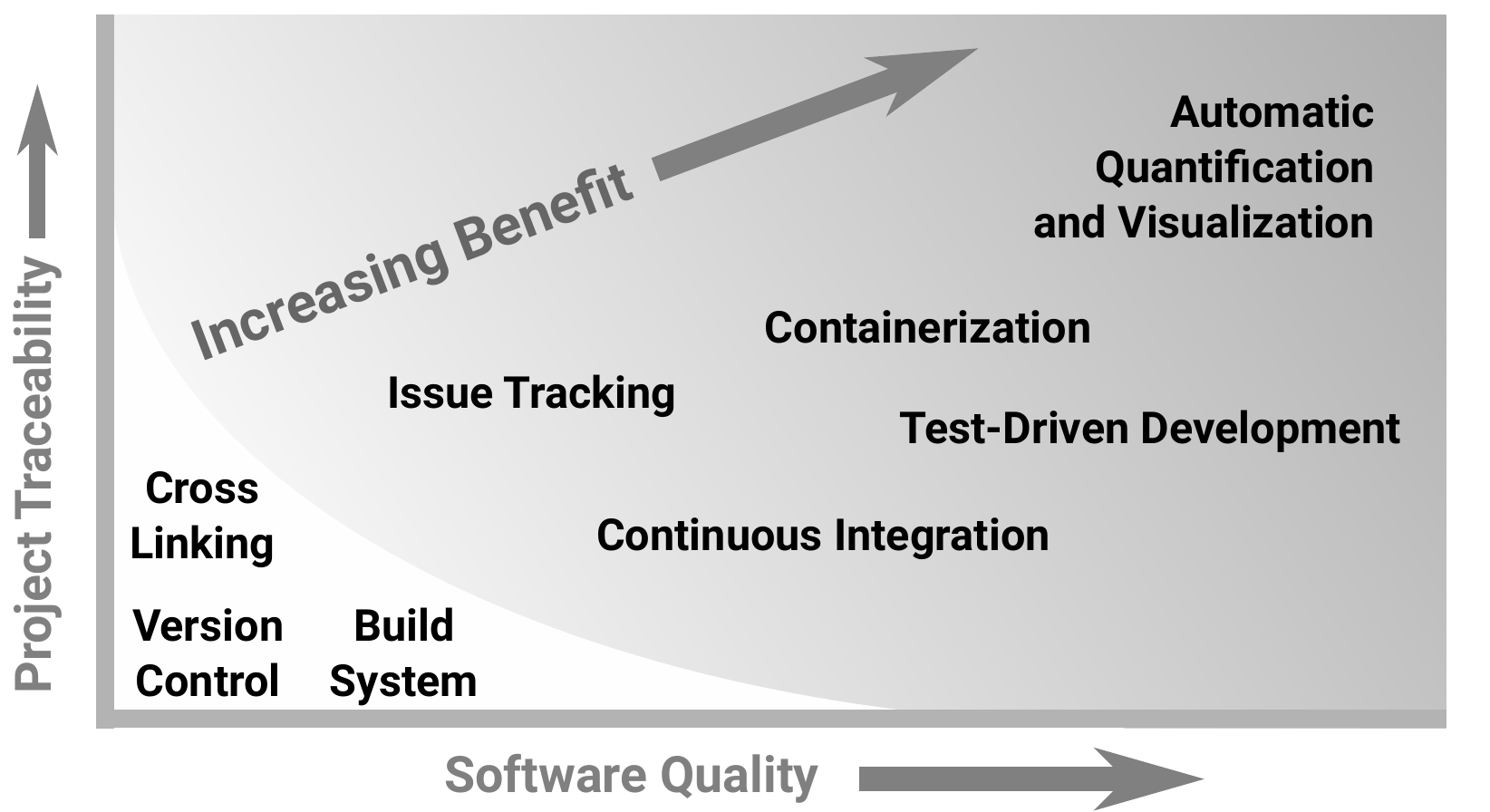}
    \caption{Overview of the proposed workflow and the technologies used. Elements in the lower white area show the \minwf{} steps. The elements along the gradient are recommended elements that comprise the \fuwf{}.}
    \label{fig:workflow-overview}
\end{figure}

An overview of our workflow's components is shown in \cref{fig:workflow-overview}.
The schematic highlights the aspects that we consider as the minimum workflow, while pointing out useful additions.
Projects should establish the minimum aspects first, and then increase towards the full workflow, as suggested in~\citep{2020:heroux:psip}.
This keeps initial investment manageable and introduces only minimal overhead to the general organization of the project.

At the very least, we recommend to
\begin{enumi}
\item use a \emph{version control system~(VCS)} for all source files of the software, as well as text files that hold experiment configurations,
\item use an established cross-platform \emph{build system} for the software artifacts, and,
\item perform \emph{cross linking} of the code, the scientific publications and data using the mechanisms provided by the VCS and persistent identifiers~(PID), e.g., digital object identifiers~(DOI).
\end{enumi}

Additionally, the project can include \emph{issue tracking} as an improvement, followed by \emph{continuous integration} and \emph{test-driven development} (TDD).
As particularly beneficial we found using \emph{containerization} for both
testing purposes and for reproducible computational experiments.
Finally, the introduction of \emph{automatic quantification, visualization and evaluation on HPC systems} proved useful to early detect any degradation in performance or numerical quality.
The latter benefits significantly from a working continuous integration pipeline being available.

\dg{The workflow described in the remainder of this paper is built on top} of available and widely adopted open source software, and we \dg{refer} to the respective software documentation \dg{wherever appropriate}. The basic workflow is described first: \cref{ssec:vcs} covers version control systems, \cref{ssec:build-system} the use of the build system, and \cref{ssec:cross-linking} covers cross linking of publication artifacts.
Thereafter, additional components of the workflow are described: issue tracking~(\cref{ssec:issue}), continuous integration~(\cref{ssec:continuous-integration}), containerization~(\cref{ssec:containerization}), test-driven development~(\cref{ssec:tdd}), and automatic quantification and evaluation using HPC systems~(\cref{ssec:tests}).

%
\section{Minimal workflow}
\label{sec:minimal-workflow}

\subsection{Version control branching-model for research software}
\label{ssec:vcs}

\begin{figure*}[htb]
    \centering
    \includegraphics[width=0.95\linewidth]{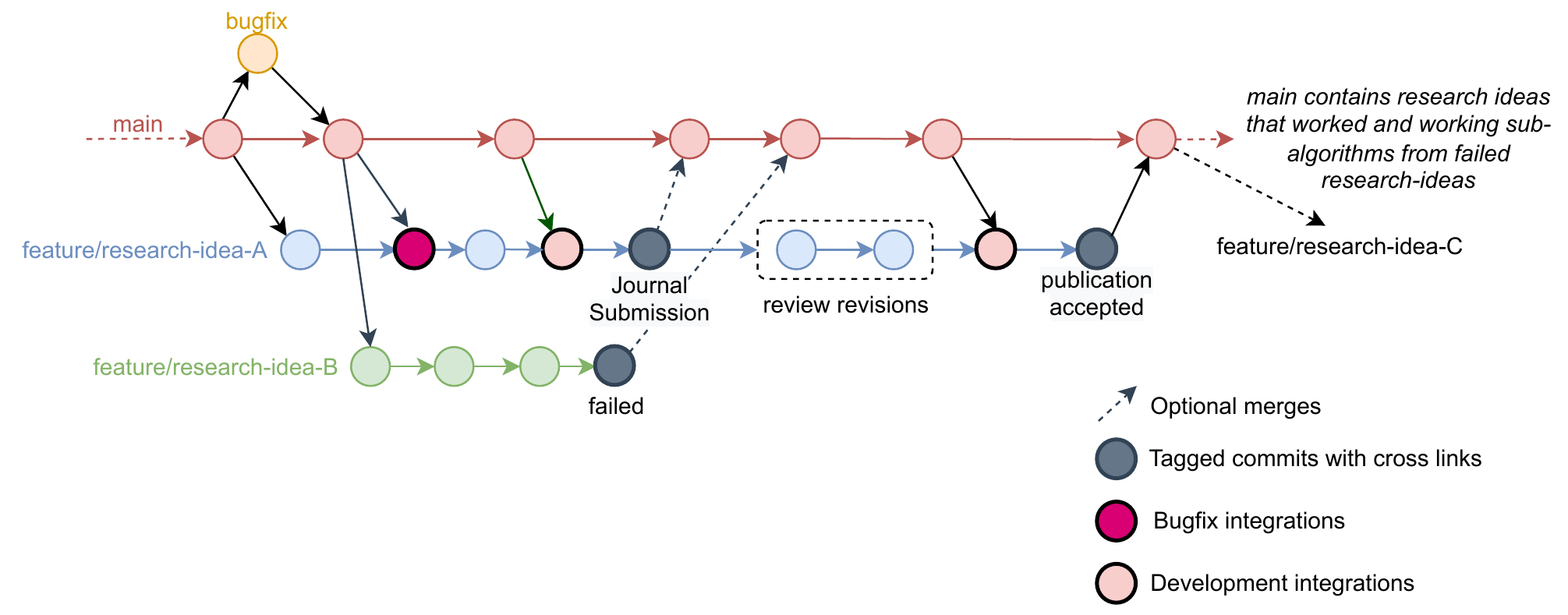}
    \caption{Schematic of the relationship between the feature-based git branching model, the peer-review process, and the cross-linking of digital research artifacts (cf. \cref{ssec:cross-linking}).}
    \label{fig:branching}
\end{figure*}

Without version-controlled software, research efforts are doomed to be repeated, which hinders progress and wastes funding resources.
Additionally, the transfer of the reasoning behind software design decisions is a challenge when the various versions diverge, and \tm{non-permanent research personnel} leaves the university.
The situation becomes particularly \dg{critical} whenever new Ph.D. students continue the work of a former Ph.D. student.

The proposed workflow approaches the challenges \tm{posed by} (re-)in\-te\-gra\-tion \tm{of different research ideas} by adopting a branching model in a version control system that \tm{combines version control with research milestones and the scientific publication process, to finally cross-link source code, research reports and result data, thus ensuring all digital research artifacts related to a scientific publication or a research idea are easily found}.
A commonly used \tm{version control system} (VCS) is \emph{git}~\citep{Chacon2014}, and many available web interfaces, e.g., Gitlab~\citep{GitLab}, Github~\citep{GitHub}, or Bitbucket~\citep{Bitbucket}, simplify its use.
For successful \tm{integration of versions}, it is crucial to follow a so-called \emph{branching model}, e.g., Gitflow~\citep{Gitflow} or Github flow~\citep{Githubflow}, to benefit most from git's capabilities.


A branching model describes how an individual or a team manage different versions. 
\dg{With a focus on small university research teams, we suggest to use the simplest possible feature-based branching model while connecting it} \tm{with the scientific publication process} \dg{to simplify result reproduction,}\tm{schematically shown in \cref{fig:branching} for an example research project.}
A research idea \dg{may require a number of different new features, which are possibly developed in parallel and on different branches
by different members of the research institution}.
\tm{We do not show research-idea sub-features in \cref{fig:branching} for the sake of clarity.}
We \dg{suggest to create} one new branch for each research idea that will contain everything \dg{that is} needed, from the perspective of the research software, for a scientific publication.
\dg{All newly developed features are merged into the research idea branch, and the research idea is finished once the results it delivers are of high
enough quality for a scientific publication, or, if the idea fails.}
\tm{The software quality - in terms of the quality of research output - is therefore connected to the quality assurance given by the peer-review process in our workflow.}
\tm{Whether an idea worked or not}, it is beneficial to write up the work carried out in a report, explaining
future researchers what has been done, \dg{what worked and what did not}.
\tm{This way, the generally completely neglected, but extremely valuable, negative scientific results are documented, at least in internal, and much better, in publicly available reports.}
Another reason for an internal report would be when a researcher leaves the institution and there is unfinished
work that may be \tm{continued} by someone else.

\dg{Before writing up the report, any new developments in the main branch should be incorporated
into the research idea branch, and it should be verified that features that had been already present 
in the research software still deliver the same results. If not, it must be investigated if the
new behaviour is an improvement or} \tm{if errors were introduced.}
Ensuring that existing features deliver results of similar quality when integrating new features can be done manually; however, in CSE context this \dg{typically takes many person-hours, and thus}, using automatic testing and continuous integration (see~\cref{ssec:continuous-integration}) is much more efficient.

\dg{At this stage, the research idea branch is mature enough to produce the results for a report,
for instance, for submission to a scientific journal. We suggest to publish both the report
and the data discussed in it in order to receive persistent identifiers (PID) that can be used for
cross-linking (see~\cref{ssec:cross-linking}).
The software itself, in the state of the research idea branch that was used to produce the
published results, should now also be published on a data repository such as
\eg Zenodo \citep{Zenodo}, or TUDatalib \citep{TUdatalib} at TU Darmstadt. This yields another PID for the software.
Next, a new commit is created on the research idea branch, which contains
only changes to the \textit{readme} file of the repository, adding
to it the PIDs of the software, data and the report publications. This makes sure that any
published digital asset produced with the software is documented. Finally, this commit is
\textit{tagged}, following a naming convention established in the group.
For instance, \textit{tags} may contain the} \tm{abbreviation of the research idea}, \dg{the abbreviation of the
journal to which the preprint was submitted, and the state of the review process
(\eg submission, accepted, revision-1, revision-2, etc.). More details on this and the cross-linking
are discussed in~\cref{ssec:cross-linking}.
}

At this point, the research team may decide if they consider the results good enough for the research idea branch to be merged with the development branch.
\dg{If preexisting features have been further developed on the research idea branch, we strongly suggest to merge it such that future unrelated research ideas can 
benefit from these improvements.}
\dg{In either case, this branching model allows the next person to comprehend the developments on the research idea branch, and possibly, start developing a new research idea based
on the one described in the report.}

The feature branch model is used further in the same way during the peer-review process \dg{of a manuscript}. 
Review comments \dg{may} lead to new feature branches and \dg{bug fixes, which are all merged before producing new results.}
Once the new version of the manuscript becomes satisfactory, \dg{the process described above is repeated for this revision: 
an archive of the research software is uploaded as a new version of the existing data item on the data repository
and cross-linked with the publication as outlined in \cref{ssec:cross-linking},
and the branch is tagged again according to the tag naming convention, suffixed with the revision number and maybe information on relevant improvements.}
Once the publication is accepted, the research idea branch is merged into the main branch,
making sure that it contains all successful ideas from published peer-reviewed work.

We want to emphasize again that we recommend this workflow also for research ideas that failed, that is, publishing code, data and a report and link them to a specific tag
in the software repository. The team may decide to keep the code with the failed research idea as a separate branch or merge it into the development branch in order to
maintain these pieces of code, for instance, if it is likely that this work will be picked up again in near future.


The major (tagged) milestones in this git branching model therefore revolve around scientific publications - digital artifacts that matter most in university research. 
Of course, this does not mean that more granular information is not available: the team can generate additional git tags for developments they deem relevant.
The core idea is to utilise git tags from major publication milestones to increase transparency and help to ensure reproducibility of results for these publications. 

Finally, here are some general recommendations for using a VCS that are and should be fundamental practice:
\emph{Write telling messages} when you commit changes, such that your project partners understand what has changed and why it was introduced.
\emph{Prefer small changes} per commit instead of minimizing the number
of commits. Each commit should introduce a single coherent change, which also makes
it easier to enrich that change with a descriptive commit message.
\emph{Avoid putting multiple features into one merge request}. If a merge request
introduces one single feature, then it is easy to give it a descriptive name.
Moreover, if the feature is developed in small and well-named commits, the code
reviewing process is much facilitated, as both the intent of the merge request
as well as the steps required to make it work are obvious.
\emph{Make sure all tests pass on each commit}. This is especially important for 
code that needs to be compiled, since commits on which the code does not build impede
the usage of tools like \emph{git bisect}. In particular, \emph{make sure the tests
always pass on the development branch}. In order to achieve this, we recommend to rebase
branches before merging them into \emph{development}, making sure that all tests still pass. 
This makes sure that there have not been any changes on the
development branch that cause the new tests of the feature branch to fail after the merge.

\subsection{Build system}
\label{ssec:build-system}

\todo[inline]{PASC review: This sec on remains very abstract and vague.}

Adoption of open source software and the complexity of scientific codes often introduce a relatively large number of dependencies.
Using an established build system simplifies handling of dependencies on different platforms.
Rules can be defined in the build system that enable the same source code to be built on different systems, where the dependencies used by the scientific code are placed in different paths.
Moreover, build systems handle platform-dependent technical details and hence, reduce the time required to maintain the build process.
We strongly advocate the use of build systems instead of writing and maintaining platform-specific build instructions as e.g. GNU Makefiles. 

\subsection{Cross-linking publications, software, and datasets} 
\label{ssec:cross-linking}

\todo[inline]{PASC review: - Findability: also for researchers who are not yet aware of these data}

\todo[inline]{PASC review: - PIDs for Code: Github/Zenodo or GitLab/Zenodo integratioon seem worthwile to men on}

\todo[inline]{PASC review: I am not convinced that "submission" is the optimal timepoint to integrate feature branches into the main branch. This should be done before submission.}

\todo[inline]{PASC review: extend the publication from arXiv to peer-review and demonstrate this with examples.}

\todo[inline]{PASC review: cover the publication process better and cover the scientific workflow - basically we did this by including secondary data into the publication. Although the publication is not a digital artefact generated by scientific software, results of the publication are, without them, there is no publication.}

\todo[inline]{PASC review: Highlighting differences between software and data repositories.}

This section covers one of the most important aspects of our proposed workflow: the cross-linking of the publication, the software and the data sets, i.e., connecting all digital assets of a particular publication.
We motivate this step from multiple perspectives.

First, the authors of the publication and their respective research groups need easy ways to retrieve all assets for a given publication, e.g., to continue their work.
This may happen at a significantly later point than the publication date, for example, when a previous idea is investigated further, or new funding is obtained.
Since research team members at universities are usually Ph.D. students and postdocs that leave the group every 1-5 years, cross-linking digital assets of a publication is crucial for sustainability.

Second, researchers who want to compare their own work to published results, need an easy way to determine all parameters used to obtain the results.
Scientific publications in CSE generally do not contain all metadata required to reproduce the results. 
Often, only the parameters relevant to the parameter variation are published, for instance, different fluid densities and viscosities, or parameters of numerical schemes investigated in the publication. 
All other parameters used by the research software that are not part of the parameter variation are not mentioned. 
However, these unmentioned parameters may be completely different in another CSE research software. 
Concretely, a publication may contain the parameters and information about approximating curvature for multiphase flow simulations, and no information about the solution of the multiphase pressure equation, since it is not related to the paper's topic. 
Reproducing the effect of the curvature model in another software on the numerical stability of the simulation, however, does require the metadata about the pressure equation and its discretization.
Therefore, archiving and cross-linking input data (input metadata) with a publication significantly increases reproducibility. 

Third, funding agencies are starting to require research to be Findable, Accessible, Interoperable, and Reproducible \citep{FAIR} for a certain number of years.
Hence, being able to retrieve snapshots and container images of software versions, experiment / simulation configurations, the input and result data sets, follows the FAIR principles closely~\citep{FAIR}.
This approach is especially applicable in CSE, where, compared to experiments, every step in the data processing pipeline is digital.

\begin{figure}[tb]
    \centering
    \includegraphics[width=0.6\linewidth]{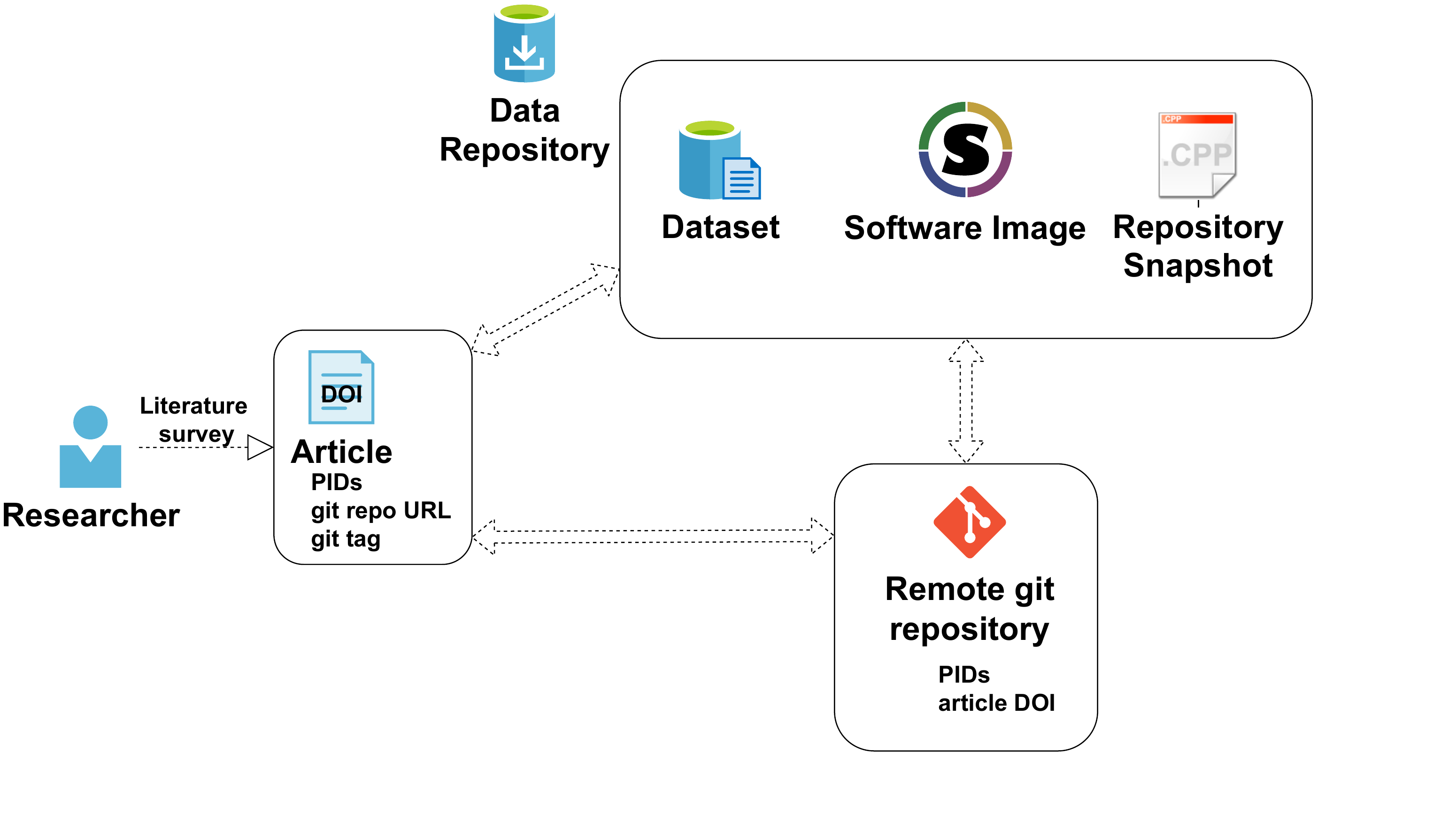}
    \caption{Cross-linked publication, data and source code using Persistent Identifiers.}
    \label{fig:cross}
\end{figure}

\paragraph{Effect of cross-linking on the publications}
If a publication serves as documentation of a method implemented in scientific software, the two should be cross-linked.
Extending the cross-linking to also include the data reported in the publication in the form of diagrams and tables (\emph{secondary data}) is the logical next step - especially considering \dg{their typically small size}.
Without archiving and cross-linking secondary data, other researchers still nowadays resort to scanning diagrams from publications, which is especially prone to error when logarithmic scales are used.
From the point of view of the research group: once the researcher responsible for the results, and the part of the source code used to generate them, \tm{leaves} the group, it quickly becomes impossible to associate the source code and the datasets used in the publication. \todo{@TM: results (, data,) and the part of source code...?}
Even if all digital assets are archived locally at the research institution the problem still persists: over the years, many publications, code versions, and datasets are generated, so the problem of findability (F in FAIR~\citep{FAIR}) grows with time, if it is not technically handled. 
From the point of view of an external researcher: the cross-linking makes it possible to retrieve all (configuration and result) data needed for a direct comparison of methods for the same simulation problem (R in FAIR~\citep{FAIR}).

\paragraph{Effect of cross-linking on the data}
The lack of cross-linking severely complicates data re-use.
To draw new conclusions or derive new results (models) from existing data, detailed information about the method that generated the data is necessary: information usually available in a scientific publication and simulation input metadata.
A new application may make it necessary to apply a method beyond the parameter set used to generate the existing data: in this case, knowing acceptable parameter ranges and finding the specific version of the research software is again necessary.

\paragraph{Effect of cross-linking on the software}
From the perspective of sustainable scientific software development and increasing scientific software quality, the issues are the same for the software, like those mentioned above for the publication and the datasets. 
For example, having only a specific version of the scientific software available, checking the results becomes problematic if the result datasets need to be re-computed.
\dg{This is especially difficult if} this involves extensive simulations that require considerable computational resources and know-how in preparing parameter variations, performing simulations, and post-processing results.
Scientific methods are improved sequentially: every research step improves on existing methodology. 
These improvements often change only some sub-algorithms of an existing method.
Thus, introduced differences are difficult to discern in the scientific code, since even relatively small research software contains many relatively complex sub-algorithms and, moreover, grows in time.
Knowing which method is implemented by the specific version of the software becomes relevant as soon as we aim to compare two methods.

\begin{figure}[t]
    \centering
    \includegraphics[clip=true,trim={0.8cm 1.8cm 0cm 17.5cm}, width=0.6\linewidth]{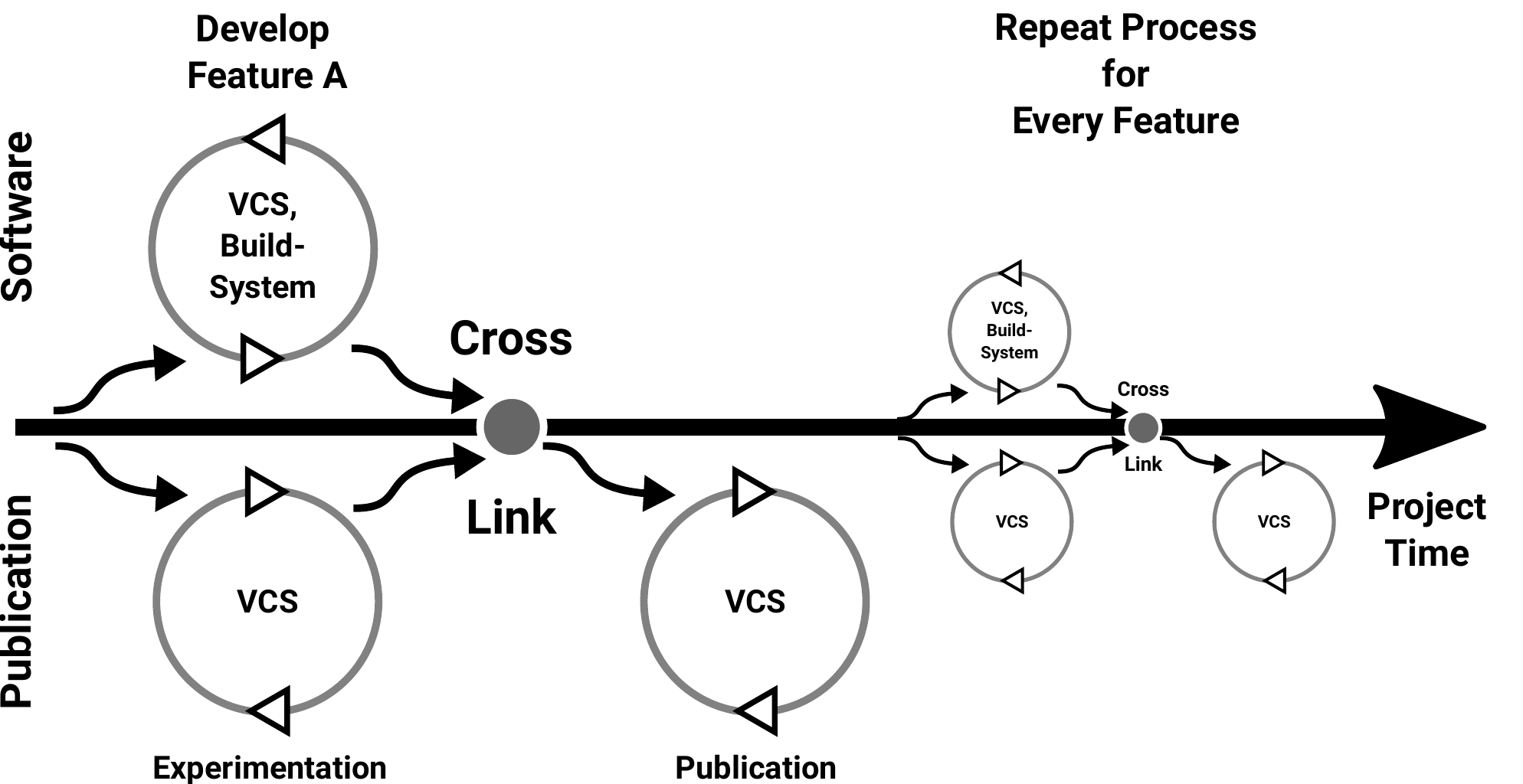}
    \caption{Schematic of a (software) project timeline and the proposed workflow steps. Software feature and experimentation is developed iteratively until the point of satisfying results. Then the publication about the method and its results is prepared and finalized. Thereafter, the next feature is approached.}
    \label{fig:workflow-timeline}
\end{figure}

The \minwf{} cross-linking connects all digital assets related to a scientific publication, excluding full simulation results and software containers, and it is very straightforward to apply.
This is important, since an overly complex workflow cannot be easily adopted with limited resources in university research groups.
A straightforward cross-linking of publications, scientific software, and result data can be achieved using Persistent Identifiers~(PIDs), as shown in \cref{fig:cross}.
%
%
As outlined in~\cref{ssec:vcs}, the cross-linking using PIDs starts at the point in the development when a report is written, that is, a preprint to be submitted to a scientific
journal, a technical report or an internal report about a failed research idea.
In \cref{fig:workflow-timeline} we visualize a timeline of a project and at which points the different parts of the workflow are placed.
For the remainder of the paper, we use the term \emph{submission} to refer to publishing a preprint, submitting to a peer-review process, or preparing an internal technical report.

\todo[inline]{@Dennis: I rewrote the remainder of this paragraph while keeping the old version as comment. Please double-check. }
At the time of submission, all feature and bugfix branches required for the research idea have been merged into the research idea branch (see~\cref{ssec:vcs}).
This branch now contains the state of the code with which the results that appear in the report are produced, and it is identified uniquely by its current commit id.
At this point, the data produced by running the code in this state are available and are used to write up the report.
After finishing the report, a snapshot of the code and the data are published into two different datasets of a data repository as \eg 
TUdatalib\footnote{TUdatalib is the data repository service provided at TU Darmstadt}~\citep{TUdatalib} or a similar service.
The data repository creates a persistent identifier~(PID) (e.g. a DOI) for every uploaded digital asset, which can now be mentioned in the report such that readers can find
the associated data and software. Now, the report itself can be published to a manuscript repository (e.g. ArXiv~\citep{ArXiv}), which further produces a PID for the publication.
If the report corresponds to a research idea that has failed, data repositories at universities usually make it possible to publish the report internally and still obtain a PID
for internal use. 

At this point there are PIDs for the code, data and the report, but so far only the report contains cross-links to the other assets. The next step is to mention the report in the
descriptions of the datasets for code and data, and place the PID of the report in the metadata. Afterwards, a commit and git tag are created in the code repository (see~\cref{ssec:vcs}),
which add the three PIDs that were obtained up to now to its documentation. Afterwards, this git tag can also be mentioned in the metadata of all three publications.
This makes it possible to retrieve a snapshot of the source code, but also to locate the version control repository and check out the tag, and continue researching.
Note that git tags are not PIDs and can be changed by the developers, so tags do not meet the requirements typically posed on long-term availability.
This three-way cross-linking \minwf{} is depicted in the first part of \cref{fig:workflow-timeline} and enables the convenient retrieval of all digital assets.
The resulting cross-linked assets are shown in \cref{fig:cross}.

One of the differences between the minimum and the full workflow cross-linking is the type of data that is published. To clarify: We separate the scientific data into
\emph{primary data} and \emph{secondary data}. Primary data are large: in the context of CSE they are full simulation results.
Secondary data are small datasets that are visualized in the publication as diagrams or tables and are instrumental in interpreting the results of the method implemented
in the scientific software and described by the publication. \emph{In the \minwf{}, only the secondary data are considered.}

To facilitate the processing of the published secondary data, scripts that produce the figures and tables shown in the publication may be added to the data publication.
In this context, Jupyter notebooks~\citep{Jupyter} have emerged as a convenient way of processing data and documenting data workflows.
In our workflow, the data generated through simulation is processed in Jupyter notebooks~\citep{Jupyter} (see \cref{ssec:tests}).

The small secondary data (e.g. CSV files in \cref{fig:testvis}) are archived together with the Jupyter notebooks and their HTML exports, the tabular information of the parameter study that connects case IDs with parameters, and the simulation input files.
In other words, the secondary data archive contains the complete directory structure from \cref{fig:testvis}, including all parameter studies reported in a publication, but without the large simulation results. 
The tabular information of the parameter study should be saved in a text format (YAML, JSON, CSV, or similar) that can be easily read by a computer program (I in FAIR~\citep{FAIR}), but also easily read by a researcher trying to find the ID of a simulation from a parameter study, that was simulated using a specific set of parameters. 
This archive makes it possible for a researcher to quickly find a CSV file (F in FAIR~\citep{FAIR}) that stores the data used to visualize a specific diagram in a publication.
The secondary data archive generally has a size below $1$GB and can easily be uploaded to and downloaded from the data repository.

It is important to note that although relatively small in size compared to full simulation results, the datasets that belong to secondary data contain all the information relevant for judging the quality of the scientific results, comparing different methods, or performing regression tests (ensuring all existing tests still pass when adding new features).

At this point, the publication may enter the peer-review process that results in reviewer comments. 
The application of reviewers' comments and further research of the method lead to the repetition of the above described process. 
If the new feature is researched and works well, it is integrated into the research idea branch, a new git tag is generated (see~\cref{ssec:vcs}), the source code archive and the
secondary data are updated as new versions on the data repository, the pre-print is modified accordingly with new cross-links, a new version of the preprint is submitted to \eg
ArXiv, and the metadata on the data repository is updated.

Although listing \tm{the cross-linking steps} here may make the workflow seem complicated, applying \tm{the steps actually} takes at most a couple of work hours, \tm{maximally a workday for an individual researcher}, \tm{and} even less in a team effort. 
Compared to the number of person-months (sometimes person-years) generally necessary to investigate a research idea, half of a workday used to follow the FAIR principles and
increase significantly the quality of \tm{research results}\todo{Not just software} is an excellent investment.

%
\section{Full workflow}
\label{sec:full-workflow}

\subsection{Issue tracking}
\label{ssec:issue}





There are two important aspects of research software development at universities that can be improved significantly by issue tracking.

First, the research software developed at university research groups is mainly developed by personnel that has non-permanent positions and leave after $1-5$ years, e.g., Ph.D. students or Postdocs.
Often, virtually no overlap period is available: the person responsible for a research direction often leaves before the new person arrives.
In addition, (undergraduate) students work on projects on a short-term basis.
This high turnover rate at university research groups, combined with contributions from different research directions and experience levels, demands documentation of the project status. 
Without project status documentation, repeating the predecessor's steps, discovering unaddressed bugs again, and the inability to understand the reasoning behind decisions already made in the project when inheriting an existing project all lead to an enormous waste of time and resources. 

Second, once the software is published, the scientific community is likely to find bugs or alternative applications for the software.
Simplifying bug reports or feature requests from the outside world is crucial for increasing research software quality.
Hence, keeping track of the status of the research software is utterly important, to allow the transfer of knowledge about open issues and potential limitations.

An issue tracking system helps in communicating such aspects clearly to new project members and people generally interested in the software project.
Issue tracking systems are typically freely available on web-based git services~\citep{GitLab,GitHub,Bitbucket}, and allow to track the status of the project.
Issues such as bugs, ideas, research cooperations, and peer-review comments are modeled as cards. 
The issue cards can be extended with links to the source code and input data and attachments (e.g., images from simulations),  and they support the chat functionality - placing the discussion about an issue in the issue itself such that it is easily findable and understandable.

Defects are reported as bug reports, and ideas are maintained as feature requests.
Moreover, ideas can be broken down into smaller sub-tasks, simplifying the onboarding process.

A straightforward way to present and update issues is the well-known Kanban approach \citep{Sugimori1977}.
It consists of a board where the issues are moved between three columns: \emph{To Do}, \emph{In Progress} and \emph{Done}.
The issues can be labelled, e.g., bug, improvement, or others, and they can be grouped into project milestones. 
If actively maintained, the straightforward Kanban issue tracking simplifies the transition of research personnel, as the status of the project, including its current milestones, can easily be understood on the visual Kanban board.
This lightweight issue tracking is already sufficient for research software developed within individual Ph.D. / PostDoc projects.
More advanced approaches for collaborative projects that involve many projects and researchers may require a dedicated issue tracking software and more complex issue tracking workflows.

\subsection{Test-Driven Development for Research Software}
\label{ssec:tdd}

\begin{figure*}[tb]
    \centering
    \includegraphics[width=0.95\textwidth]{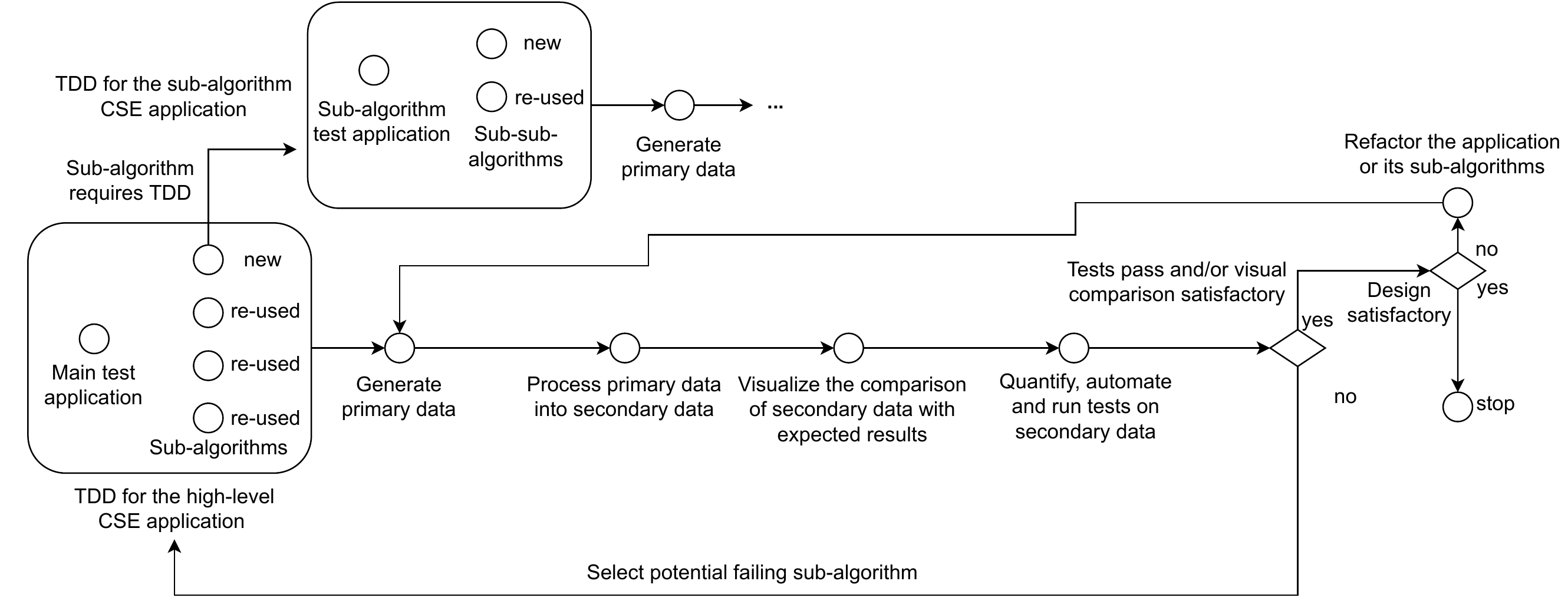}
    \caption{Test-Driven Development (TDD) applied to research software in resource-limited university research setting. TDD is applied in a top-down approach, starting with the CSE application that should generate research results, then applying TDD to its sub-algorithms recursively, on a per-need basis.}
    \label{fig:rsetdd}
\end{figure*}

\todo[inline]{This section is meant to be a substitute for the two following sections
whose text I haven't deleted yet, it is just commented out in the .tex file. One thing I left out are the smoke and 
production tests - I thought they are not fundamental thing of TDD? We should
maybe discuss that.}

Test-Driven Development (TDD, \citep{Beck2003}) is a practice of writing test applications (programs) first, \tm{before implementing the algorithms that are supposed to deliver the test results.} 
This forces \tm{programmers} to think about the code from
the perspective of the user, which typically yields a cleaner API \tm{(Application
Programming Interface) - the kind of API, the programmers would like to use themselves.} 

TDD consists of \tm{three phases}: red, green and refactor.
\tm{The first, \emph{red}, phase begins with the implementation of the test applications, and,
as no algorithmic implementation is available to perform the actual computation, the tests fail.}
The tests define the return types and arguments of algorithms used in the tests, thus defining the API of the algorithm library.
\tm{The algorithm implementation follows, and it is modified until the tests pass, to enter the \emph{green} phase of TDD.}
\tm{In the final, \emph{refactor} phase,} the passing tests now allow the developers to refactor the implementation while using the tests to verify that the algorithmic implementation still works as intended.
\tm{Refactoring is
done until a modular (re-usable and extensible) software design is achieved.}
Then, the next algorithm is approached, by starting over with writing the tests first.


We propose an adaptation of the TDD approach to research software in CSE. Our top-down TDD for CSE research software is outlined in \cref{fig:rsetdd}. 
Numerical methods in CSE consist of different complex
sub-algorithms.
Some methods are based on rigorous mathematical theory and have provable properties, e.g., convergence, or error estimates, others are based on approximations whose errors cannot be theoretically estimated.
In both cases, extensive and automatic verification and validation are crucial, since proven properties are not equivalent to a correct implementation.
Moreover, if the implementation does not produce the expected results, it is necessary to determine which part of the method's implementation causes the
problem, which is generally done by testing.
\tm{If TDD is applied for the development of research software with the goal of producing the (test)} results for a scientific publication, \tm{the focus is shifted from attempting to implement low-level unit-tests of all algorithms, to preparing only those verification and validation tests, that are relevant for the subsequent scientific publication}.
\tm{This way, tests that are written first, without providing the algorithmic implementation, are very high-level CSE applications, for example: partial differential equation (PDE) solvers.
Existing publications, that are supposed to be outperformed, define the test results. 
Alternatively, the results are defined by experimental data, or by the method of manufactured solutions.
Without a single line of algorithmic implementation, it is therefore possible to define the type of primary data generated by the CSE high-level test application  (e.g., velocity, pressure, and temperature fields), and what is expected of them (macroscopic quantities, e.g. drag resistance), as shown in \cref{fig:rsetdd}. 
Furthermore, it is possible to visualize the not-yet-available (empty) results against expected results - on purpose generating an artificially failing comparison.}

\tm{Once the research tests are prepared, the algorithmic implementation follows for the high-level application, by re-using algorithms from legacy research software, or, if necessary, implementing algorithms from scratch.
For algorithms implemented from scratch, the RSE TDD approach branches off and cycles through its phases until the algorithms implemented from scratch are refactored, in a recursive top-down approach, since they also consist of sub-algorithms.
Re-used sub-algorithms from legacy code on every level are assumed to be working correctly, unless the TDD testing shows otherwise, in which case suspicious sub-algorithms are selected for further TDD, as shown in \cref{fig:rsetdd}.
When the high-level application is not generating correct results after completed algorithmic implementation, TDD is applied to its sub-algorithm that is selected to be the most likely cause of the problem during result visualization/analysis.
In CSE, algorithms are generally very complex because they model complex physical phenomena. 
For example, a PDE discretization that is unit-tested for a set of Initial and Boundary Value Problems (IBVP), does not necessarily work for a IBV problem outside of the tested set.
Automatic testing and visualization of results in the red phase help the programmers more easily isolate troublesome sub-algorithms that require more TDD.
This way, finalizing the green phase for the high-level application does not require unit-testing to be applied on every sub-algorithm of a CSE application, which is not possible to do in a university research setting because of the lack of dedicated resources. 
Only those sub-algorithms that are not correct in the context of expected research results are further developed with TDD.}
\tm{ As work continues on different scientific publications, the automatic test suite grows with new tests being added for new publications and the overall test coverage increases along the research roadmap, with working and useful sub-algorithms from both failing and succeeding research ideas integrated into the main software version following the RSE version-control branching model \cref{ssec:vcs}. 
}

\subsubsection{Test qantification and visualization}
\label{ssec:tests}

\begin{figure*}[tb]
    \centering
    \includegraphics[width=0.95\textwidth]{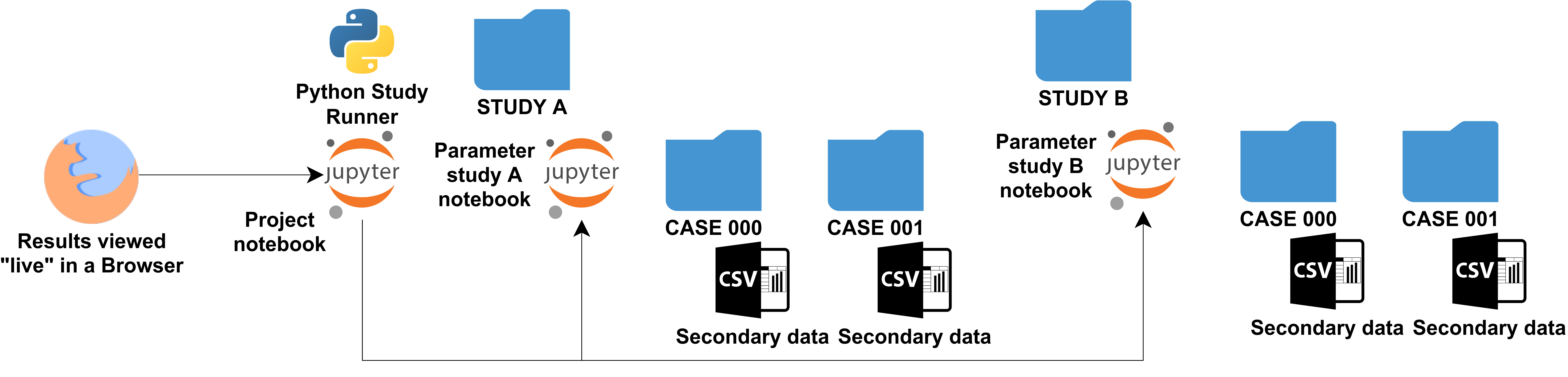}
    \caption{Different parameter studies are orchestrated by a Study Runner. All studies are implemented using Jupyter notebooks, which contain a problem description and the visualization of results. This allows for \emph{live} inspection of test result data. Structured metadata is exported into files to be available for subsequent use in publications and data cross-linking.}
    \label{fig:testvis}
\end{figure*}

Testing scientific software involves running parameter variations, so-called studies, as shown in \cref{fig:testvis}.
They often include hundreds of simulations, so called cases, \emph{cf.}~\cref{fig:testvis}.
The large number of test cases makes it important to be able to quickly identify failing tests and re-run them after changing the algorithm implementation to fix the problem.
Hence, the respective data, input and output, needs to be well organized.

The main input of such parameter studies are simulation parameters that are varied across the components of a \emph{parameter vector}.
This is, typically, performed automatically by a \emph{study runner} on an available HPC machine, as the production tests consume a significant amount of resources to execute.
In addition to running the respective case, the study runner also builds a mapping, using unique IDs, from each simulation case to the particular input parameter vector used.
Moreover, it exports structured metadata \tm{stored in both human and machine-readable standardized and open file format}, that can be used subsequently as part of the cross-linking process, \emph{cf.} \cref{ssec:cross-linking}.
\cref{fig:testvis} shows a possible organization of the data in folders with a naming scheme of simple ascending integers used as the unique IDs.

Being able to examine test results of the parameter studies run as early as possible is crucial to catch problematic executions timely, i.e., stop their execution, thus, limit the \tm{wasting of compute resources, reduce the negative impact on the priority on the HPC machine, and speeding up the research process.} 
Being able to perform real-time analysis of test results is, thus, fundamental.
Manual inspection of tables and diagrams is \tm{a natural part of the research process: it is often not possible or necessary, and sometimes not tractable, to automatically quantify the comparison of secondary data, resulting in an automatic pass/fail status of the test.}
A straightforward solution to the challenge of (almost) real-time manual test result analysis is the use of Jupyter notebooks~\citep{Jupyter}. 
A Jupyter notebook is written for each individual study, \emph{cf.} \cref{fig:testvis}.
The project notebook, i.e., the one executing the study runner, is created to link all the other notebooks.
This effectively links together all the tests used to verify / validate the scientific software in a single place.

Each Jupyter notebook is used to document its parameter study.
\tm{The test documentation covers the mathematical description of the CSE problem that is solved by the test.}
Such descriptions include which equations are solved, the initial and the boundary conditions, as well as physical parameters that are investigated.
The Jupyter notebooks then process secondary data generated by the simulation (or a post-processing utility that runs alongside the simulation), and store them as ASCII CSV files in \cref{fig:testvis}.
The central project notebook can be viewed \emph{live} locally in a browser, after the Jupyter server is started remotely on the HPC cluster.
This straightforward use of Jupyter notebooks makes it possible to check simulation results of many studies / cases that may run for a long time on a cluster, and stop those early that are not generating satisfactory results.
\tm{Additionally, since Jupyter notebooks are used for sequential processing of secondary data they use as input, they are always executed anew, and do not require information on the sequence of execution of their cells (provenance).}

An important benefit of this visualization process is that diagrams and related secondary data processed by the notebooks are used directly in the scientific publication.
They are stored by the notebooks in dedicated folders, that can be easily synchronized with the publication folder, as soon as the results are satisfactory.
Moreover, the solution is well suited for \tm{cross-linking of digital research artifacts (cf. \cref{ssec:cross-linking}), containerization (cf. \cref{ssec:containerization}), and continous integration of research software (\cref{ssec:continuous-integration})}.

A Ph.D. or a postdoctoral researcher focusing on a CSE methodology will not generally have the resources needed to alter the file format of primary data from a legacy CSE research software, which is not the case for secondary data.

On the other hand, \emph{secondary data is precious}: It is the basis for acceptance or rejection of scientific publications, decisions on whether milestones have been reached, and conclusions about one method outperforming all others.
There are two challenges in handling secondary data that our workflow addresses: handling metadata and interoperability. 
We propose a very straightforward solution. 
Secondary data is minuscule compared to primary data. 
Storing secondary data using an ASCII CSV format while saving metadata alongside result data directly in the columns solves both the issue of handling metadata and the interoperability of secondary data.
However, storing metadata next to data in columns introduces information repetition. 
For example, one can think of secondary data as a vector (row) uniquely identified in a parameter variation by another vector - a vector of parameters. 
In parameter variations, other parameters are held constant as one parameter varies.
Established data formats (e.g., HDF5 \citep{HDF5}) rely on a tree structure to model parameter variations without repetition of values. 
Contrary to this approach, for secondary data, we repeat non-varying parameter values alongside varying parameter values and store both as secondary-data columns. 
This way, all secondary data, and metadata are stored in a single table in ASCII CSV. 
This approach makes our secondary data completely interoperable, as any software that can process tabular data can process our secondary data.
The information about which column stores metadata information and which secondary data becomes apparent if the column names follow the nomenclature from the scientific publication. 
Furthermore, this approach trivializes the inclusion of metadata into data analysis:  using metadata for sub-set selection and other statistical or sensitivity analyses. 
Not relying on complex file formats simplifies the processing of secondary data immensely and saves time, with a negligible overhead caused by (meta)data duplication.
A concrete example of storing metadata together with secondary data in tabular form is shown for hyperparameter tuning of an artificial neural network (NN) in \cref{table:metadata}. 
Fixing one parameter and varying others causes data duplication, which can be ignored for secondary data because of its small size, and potentially used in further data analysis to better understand, in this case, the training behavior of the NN model. 

\begin{table*}[!htb]
    \scriptsize
    \begin{tabular}{llrrrrr}
\toprule
{} HIDDEN\_LAYERS &  OPTIMIZER\_STEP &  MAX\_ITERATIONS &  DELTA\_X &  EPOCH & TRAINING\_MSE \\
\midrule
10,10,10,10 &          0.0001 &            3000 &   0.0625 &      1 &        1.091560 \\
10,10,10,10 &          0.0001 &            3000 &   0.0625 &      2 &        1.082970 \\
10,10,10,10 &          0.0001 &            3000 &   0.0625 &      3 &        1.077200 \\
10,10,10,10 &          0.0001 &            3000 &   0.0625 &      4 &        1.072650 \\
...         &          ...    &            ...  &   ...    &    ... &        ... \\
10,10,10,10 &           0.001 &            3000 &   0.0625 &      1 &        0.992354 \\
10,10,10,10 &           0.001 &            3000 &   0.0625 &      2 &        0.991959 \\
10,10,10,10 &           0.001 &            3000 &   0.0625 &      3 &        0.995102 \\
10,10,10,10 &           0.001 &            3000 &   0.0625 &      4 &        0.996143 \\
...         &          ...    &            ...  &   ...    &    ... &        ... \\
\bottomrule
\end{tabular}

    \caption{Storing metadata (hyperparameters) together with secondary data (epoch, MSE) in tabular form.}
    \label{table:metadata}
\end{table*}

\subsection{Containerization}
\label{ssec:containerization}

\todo[inline]{PASC review: - It could be worth to men on that also the container les can be cross-linked}

The reproducibility of results is a fundamental principle of research.
In addition, being able to use software in a productive environment is highly relevant. 
However, the multitude of platforms provides a challenge, given the usually limited number of personnel in academic research groups, especially when required on top of the demanded FAIR principles~\citep{FAIR}.
An approach for cross-platform interoperability is the encapsulation of the software environment in so-called containers.
Containers, e.g., Singularity containers~\citep{Singularity, Kurtzer2017}, make it possible to execute software in any environment, given that it provides a suitable container runtime.
Hence, results can be reproduced more easily if the researcher provides containers that include the software and configuration files used for a specific publication. Large input data sets should not be included in the
container but should be published separately in a data repository. This keeps
the container size as small as possible, for instance, for users that want to
use the software on their own data set. For the reproduction of results, the
input data can be downloaded on-the-fly from the data repository.

How the container is built is specified in \emph{recipe} files.
They list all required dependencies, the software, how it is built and which data to copy into the container. However, note that copying data into the
container makes the recipe itself not comply with the FAIR
principles~\citep{FAIR} as one cannot reproduce the building process without
that data. Therefore, it is beneficial to only use data from online
resources with a guarantee of availability, for instance, from dedicated data
repositories. The same holds also for code: one should prefer to pull used sources
from release pages or "git clone" into a repository with a specific commit or tag
in the build recipe instead of copying source code into the container. In general,
git repositories have no guarantee to exist over long time periods, as they could
in principle be moved or removed. Therefore, if available, one can clone into the
mirrors hosted at \href{https://www.softwareheritage.org/}{softwareheritage.org/}
which are guaranteed to be conserved over a long time frame.

Finally, the recipe could define a set of commands that can be invoked on the container, e.g., to reproduce a specific result.



\subsection{Continuous integration of research software}
\label{ssec:continuous-integration}

\begin{figure*}[!ht]
    \centering
    \includegraphics[scale=0.5]{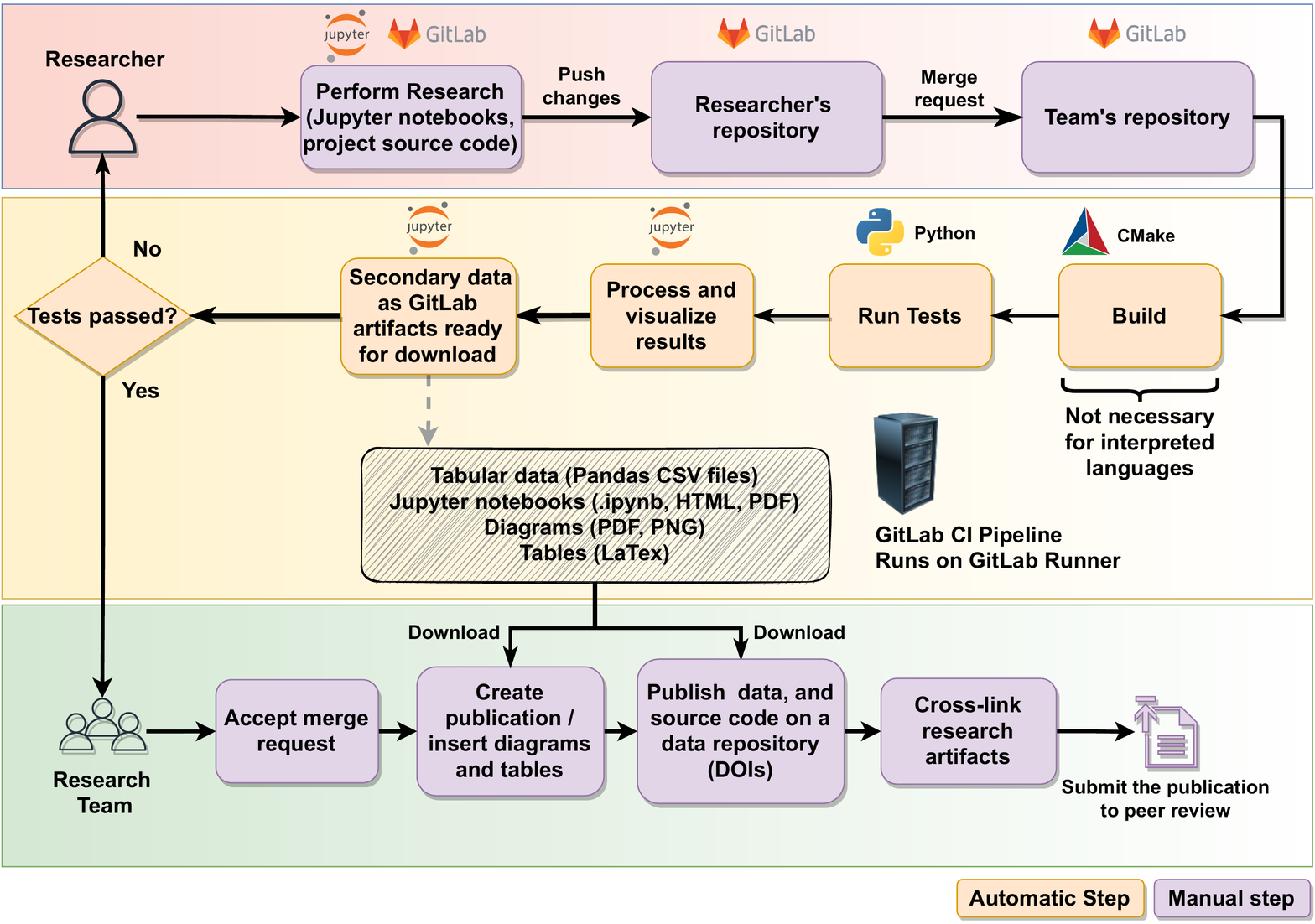}
    \caption{Continuous integration workflow: the merge request triggers the CI pipeline to execute. This builds the software, runs all tests and produces the visualizations. These are published for inspection. If tests pass, the merge request is accepted and the changes are added to the main repository. If the tests fail, the developer inspects the results and re-initiates the process.}
    \label{fig:ci_cd}
\end{figure*}

Continuous Integration~(CI)~\citep{Fowler2006} is the practice of frequent integration of changes in a shared version of the software.
In this section, a straightforward CI workflow is presented for research software, that increases its quality without introducing a significant workload overhead for the researchers.

To understand CI, and its benefits, it is beneficial to briefly reiterate the connected branching model, \emph{cf.}~\cref{ssec:vcs}.
A new feature is developed using a so-called \emph{feature branch} (research idea), while the current stable version of the software is maintained on the \emph{main branch}.
In a more traditional sense, CI recommends frequent, e.g., daily, integration from features onto the main branch.
Daily integration of changes is not tractable for scientific software, because the features are driven by scientific research: it is often not immediately clear if the developed methodology will work.
Therefore, this integration is performed when the generated results are suitable for the submission of a scientific publication to a peer-review process.
On the other hand, bug fixes and potential improvements to the main branch can be integrated into the feature branch.
To anticipate diverging individual feature branches, this integration should be done frequently, e.g., weekly.
This ensures that improvements to common, shared components are integrated into the individual feature branches.
Moreover, integrating from main to feature branch on a weekly basis, makes sure that the developments achieved in a successful project are integrated by everyone in the research group, while the responsible developer is still working in the research group.
Hence, no working software artifact is lost.

Even though the integration itself may not always be continuous, there are aspects of CI that help to increase code quality.
\cref{fig:ci_cd} shows a CI workflow example.
%
Feature development happens in the user's repository.
When the feature is implemented together with its tests, the user opens a merge request~(MR), i.e., a request to integrate their changes into the main branch.
This triggers the CI pipeline, where the build system, CMake in the example, is used to check the compilation of the software on different platforms.
Once the software passes this stage of so-called build tests, the actual tests are executed on the GitLab runner.
Besides small-scale unit tests, this involves the execution of parameter studies and populates the respective folder structure with secondary data, as shown in~\cref{fig:testvis}.

The automatic execution of tests as part of the CI pipeline is an important aspect to maintained software quality, and simplifies re-integration.
A merge request triggers the execution of the CI pipeline, and the MR should only be accepted if all tests have passed.
This ensures that the integration of all changes, i.e., feature branch and master branch, do not lead to errors in parts of the software.

We explain an important additional detail of our CI pipeline, geared towards the needs in CSE.
Before the quantification of tests causes tests to either fail or pass, results are processed that show \emph{why} the tests failed / succeeded.
Often, the textual output of the test environment is not sufficient to diagnose test failure.
Especially in the context of many CSE disciplines, test results require visualization in the form of diagrams or $2$D/$3$D visualization to determine the cause of an error.
For this purpose, the proposed CI pipeline from \cref{fig:ci_cd} contains jobs that export the Jupyter notebooks as HTML. 
\tm{The result is the complete visualization of test results, uniquely identified with a commit.} 

\todo[inline]{Proposed change by @Dennis: Besides unit tests, the test suite comprises parameter studies that involve short simulations for varying grid refinement and/or other parameter values. For simple test cases with analytical reference solutions, convergence rates upon grid refinement are determined and compared to the expected values. These convergence tests are performed for a wide range of choices of other model parameters to ensure that the implementation handles these parameters correctly. For more complex physics, for which no analytical solution can be derived, tailored test cases are used to check if physically meaningful results are obtained upon specific parameter variations. For instance, in the context of computational fluid dynamics, one can check that the magnitude of the velocity scales as expected with the density of the considered fluid. These relationships between target quantities and variations in model parameters are visualized in the test pipeline such that this information is readily available to the developers in case a failure (i.e.\ an unexpected relationship) is detected.\\
Finally, we strongly advocate the use of regression tests, in which the result produced by a short simulation in the test pipeline is compared against a result obtained by the software at an earlier stage. We recommend that for any physical process that is implemented, a small simulation promoting that process is added to the test suite together with a reference solution obtained at the point of writing the test. Upon test execution, the newly produced results are compared to the reference solutions using tolerances that are small enough to capture physically relevant deviations. While this does not guarantee physically correct solutions, it helps to detect changes in the behavior of the simulator upon any changes to the code. This means that in case a bug fix leads to failing regression tests, the reference solutions has to be updated accordingly. In order to facilitate regression testing, we have developed a Python package that allows for the detection of deviations in simulation results, supporting a variety of standard file formats (\href{https://gitlab.com/dglaeser/fieldcompare/-/tree/main/fieldcompare}{gitlab.com/dglaeser/fieldcompare}).

TODO: Actually use fieldcompare in the CI, we currently still use the old package, but I need to first implement the field filtering that we use in the CI into the new package.}

Provided all tests pass, the acceptance of the MR in \cref{fig:ci_cd} is decided by the responsible team member, who can also provide feedback on code quality.
A web interface to git, such as GitLab, simplifies the code-review process within the research group.

\todo[inline]{The HPC tests were not addressed in detail, and we don't yet have the solution for HPC reproducibility. We should discuss this.}

\subsection{Cross-linking publications, software, datasets, and software images}

While the \minwf{} cross-linking ensures the source code, the publication, and the secondary data are cross-linked, it can be extended further with primary data and containers (see \cref{ssec:containerization}). 
Publishing and cross-linking primary data enables readers of the publication to gain further insights into the results and draw new conclusions.
Adding software containers to the cross-linking workflow enables results of the scientific software to be reproduced more easily and across different computing platforms.
Here the additional steps are covered, \emph{cf.} \cref{fig:workflow-overview}, that are used to achieve these improvements.

The primary data is archived in a slightly different way than the secondary data because of its size.
Each parameter study is stored separately to enable the researcher to only download a specific parameter study, which saves time and network bandwidth.
The directory structure, shown in \cref{fig:testvis} is used similar to archiving secondary data, the only difference being that each parameter study is stored in a separate archive. 
This makes it possible for the researchers to access subsets of results, for example, "STUDY A" or "STUDY B" in \cref{fig:testvis}, with respective Jupyter notebooks, their HTML exports, and the parameter study tabular information.
Datasets that are generated for subsets of the primary data are uploaded to a data repository and their PIDs are used for cross-linking as described for the \minwf{} above.

Containers in this respect can be seen as single image files that contain everything necessary to execute user-specified commands.
These commands are defined when the container is created, and, since containers are immutable, will produce the same result when invoked.
In our workflow we use Singularity containers~\citep{Kurtzer2017}, as the technology is supported by our HPC center, and the images are single files that can easily be
uploaded to a data repository. It is important to publish the recipe for building the image alongside with the actual image, as this is an important documentation of the
requirements and dependencies of the published research software. \dg{A} problem \dg{with the} code publication of the \textit{minimum workflow is that it} only
contains the research software itself. However, that may depend on several other software packages, specific compiler versions, etc. This can make it difficult for interested
peers to get the software running on their system, \dg{even if these dependencies are explicitly stated.}
Moreover, the software may only run with specific versions of these dependencies
or possibly produces different results with other versions.
The recipe for building the image explicitly documents how the environment of the researcher who produced the results
published in the report can be instantiated and the image enables users to directly spin up a container with this environment without having to install anything.
\dg{In order for this recipe to yield the same results over time, it is important that it uses a version-pinned base image and only installs pinned versions of the
dependencies.}
\todo[inline]{@Dennis: cite paper "Johnny doesn't know how to build"?}

\section{Minimal working example}
\label{sec:minimal-example}

We have prepared a Minimal Working Example (MWE) - a template CSE research project - that can serve as basis for adopting the proposed workflow. The workflow MWE contains a numerical C++ application implemented as a version-controlled open-source project \citep{MWErepo}, that generates a small data set, visualized by a Jupyter notebook, referenced in a minimal example of a scientific report (publication) \citep{ResearchReport}.

The MWE research software is a simple C++ application that evaluates numerical derivatives of a polynomial using finite differences in a 1D interval. The finite differences are compared against exact values for a prescribed polynomial. The MWE research software is version-controlled (\cref{ssec:vcs}), and built using the CMake build system (\cref{ssec:build-system}). The example data sets, the repository snapshot, the active repository, and the minimal report are all cross-linked (\cref{ssec:cross-linking}).

The MWE is archived at the \href{https://tudatalib.ulb.tu-darmstadt.de/handle/tudatalib/3522}{\textcolor{blue}{TUdatalib data repository}}. The research data \citep{Data} and the snapshot of the research code \citep{CodeSnapshot} are archived, and cited in the example Research Report \citep{ResearchReport}, along with the URL of the "live" code repository. Once the research report \citep{ResearchReport} is uploaded to the repository, the \emph{metadata} of the other data items are updated, by adding 'dc.relation.isreferencedby' element, that denotes the data item is referenced by another item. This way the \href{https://tudatalib.ulb.tu-darmstadt.de/handle/tudatalib/3530?show=full}{\textcolor{blue}{full metadata record}} of a data item on the repository stores the cross-linking information, that can be updated as data-items evolve. For example, as the peer-review process progresses or new milestones are reached, new sub-versions of data items obtain new DOIs that denote sub-versions, eg. \url{https://doi.org/10.48328/tudatalib-921.2} for a new version of the Research Report. This makes it possible to continue updating the Report, Data, Code Snapshot with the feedback from the peer-review process or during the development of new and improved methods.

A minimal CI pipeline is configured for the GitLab \citep{GitLab} MWE source code repository \citep{MWErepo}, demonstrating the use of Jupyter notebooks for data processing and visualization. The research data linked with the report \citep{ResearchReport} is the artifact generated by the CI pipeline \citep{Data}. For realistic CSE tests, Jupyter notebooks contain detailed information about the test setup: geometry of the problem, initial and boundary conditions, model parameters.

\section{Case studies}\label{sec:case-studies}

\todo[inline]{PASC review: results are not presented in a quantifiable manner (?) - the reader is left to trust the authors. Perhaps this is solved by a template repository.}

This section presents two software projects that follow the minimum workflow and apply, in addition, parts or all of the full workflow.
We highlight which parts of the workflow were introduced at what time of the project and how it helped to improve the software or its development process.

\subsection{geophase}
\label{subsec:ccs}

The geophase library \citep{geophase} implements geometrical algorithms and models for intersecting non-convex polyhedral volumes with non-planar faces in the \cppl{} programming language. 
Those algorithms are used in \citep{Maric2020ccs} for the positioning of a piecewise linear fluid interface in an unstructured geometrical Volume-of-Fluid method for simulating two-phase flows \citep{Maric2018,Maric2020rev}. 
The development of the Consecutive Cubic Spline (CCS) interface positioning algorithm \citep{Maric2020ccs} in the geophase library serves as a case study for the proposed workflow for increasing the quality of scientific software.

\subsubsection{\Minwf{}}

The git VCS is used for version control (\cref{ssec:vcs}) and the geophase library \citep{geophase} is actively developed on TUGitlab \citep{TUGitLab}. 
To simplify cross-platform installation and handling of dependencies the CMake~\citep{CMake} build system is used (\cref{ssec:build-system}). 
For the submission of the CCS algorithm \citep{Maric2020ccs}, the cross-linking of the preprint, the source code and the resulting data was done (\cref{ssec:cross-linking}).
Applying the proposed minimal workflow in the peer-review process increases the transparency, because the reviewers and the rest of the scientific community can access the preprint, the source code and the data.

\subsubsection{\Fuwf{}}

Issue tracking (\cref{ssec:issue}) is used to track the status of the project on GitLab using Kanban boards. The Continuous Integration (CI) is used for "geophase" as described in \cref{ssec:continuous-integration}. 
As shown in \cref{fig:ci_cd}, this also includes publishing Jupyter notebooks and secondary data as CI artifacts, available for download from the GitLab/Hub web interface. 

For geophase, the results from successful CI pipelines are also publicly available \citep{geophaseCI}.
This makes it possible to quickly isolate possible negative effects between different versions, before quantifying them and testing them inside the CI pipeline.
For the submission, a Singularity \citep{Kurtzer2017} container is created (\cref{ssec:containerization}) that contains both the geophase repository and its software enviornment. 
The Singularity container is configured in a way that enables execution of different tasks with a single command:
\begin{enumi}
\item the geophase library can be cloned
\item the source code of the geophase library can be built using the dependencies stored in the container, 
\item all tests can be executed and re-started, 
\item the jupyter notebook in the container can be served from within a container to visualize results.
\end{enumi}
Currently, the generation of the Singularity container (although straightforward) is still done manually, because the delivery of the container is reserved for a submission. 
Alternatively, the creation of the Singularity container can easily be included into the CI pipeline, which would, in turn, result in Continuous Delivery across different platforms. 
The geophase library was developed using TDD (\cref{ssec:tdd}), the tests have been implemented using the GoogleTest testing framework and categorized using CTest. 

A more complex future step extends the use of the Singularity container to run performance tests on the HPC cluster as a part of the CI pipeline. 
Once the production test execution on the HPC cluster is integrated in the workflow shown in \cref{fig:ci_cd}, the next step is to automate the publishing of the data sets using the API of the data repository.
It is important to note, however, that it is not necessary to automate every aspect of the workflow, because new features are developed very slowly in scientific software, and satisfactory results are obtained sometimes after many months or years of work.

\subsection{PIRA}
\label{subsec:pira}

PIRA\footnote{PIRA: \url{https://github.com/tudasc/pira}.}\citep{2018:lehr:pira, 2019:lehr:piraII} is a performance profiler for \cl{}/\cppl{} applications.
It consists of different modules, implemented in different languages, that are maintained across multiple repositories, and various other external dependencies.
Hence, a particular version of PIRA consists of particular versions of various software projects.
When applied to a target application, the user specifies a configuration file that determines various parameters.
In particular, how the target program is executed and where to store profiling results.
This configuration file is, in its nature, similar to simulation parameters.

\subsubsection{\Minwf{}}
As suggested in the proposed workflow, the project uses the git VCS.
PIRA includes its different components as \emph{git submodules}, i.e., it tracks a specific version of theses modules.
The project is implemented in Python and some of its dependencies are implemented in \cl{}/\cppl{}.
The build system CMake\citep{CMake} was used from the start for all native code modules.
For its publications, the respective version of PIRA was tagged, which also includes the specific version of the submodules.
The configuration files are stored in a separate git repository, and a branch is created for each publication.
Cross-linking, as proposed in this paper, was not performed for the initial PIRA publications as it was still unclear which services to use.

\subsubsection{Additional workflow components in PIRA}
PIRA introduced a particular branching model later in the project.
The motivation for its introduction was twofold: the release of PIRA as open source software, and multiple students working on PIRA.
To simplify contributing, and guarantee a certain level of quality, the Gitflow~\citep{Gitflow} model was adopted.
The introduction of the Gitflow model helped with stabilizing the project's public repository significantly.

In addition to the introduction of the Gitflow model, the issue tracker and continuous integration were introduced to improve software quality further.
The issue tracker is used to report bugs and suggest new features or usability improvements.
Student assistants contribute by taking over responsibility for a certain issue.
Following the Gitflow model, they create a new feature branch, implement the feature, and re-integrate it using a merge request.
Since the adoption of the Gitflow branching model, the issue tracker and continuous integration, the public repository contained only one partly dysfunctional commit, i.e., a specific version in which some of the unit tests were failing.
The CI helps significantly with the development of features in the software.

A sometimes overlooked aspect of a working CI environment is that it serves greatly as a reference of what \emph{should} work.
Especially in a university setting, e.g., students on-boarding for a Bachelor or Master thesis, being able to point to a working configuration and environment of the software is beneficial.
The on-boarding of students, before the introduction of CI, consisted of multiple meetings to set up the environment and install required dependencies.
The introduction of CI mandated to improve this aspect.
Moreover, students start with the CI configuration, and only \emph{specific questions} of the setup are asked and resolved in personal (or virtual) meetings.

Lately, the containerization of PIRA, to simplify the CI environment, and gain first experiences with using PIRA in the cloud~\cite{2020:sokolowski:hycloud} was started and is being explored further.

\section{Conclusions}
\label{sec:conclusions}

The proposed Research Software Engineering workflow for Computational Science and Engineering significantly increases the quality of scientific results with a minimal workload overhead, which makes it applicable in a university research setting with limited dedicated resources for RSE. 
Placing the focus of the workflow on scientific publications makes the workflow attractive for researchers that already mainly develop CSE research software to generate scientific publications as their primary scientific output.

\dg{
The minimal workflow combines an established build system with a simple feature-based version-control branching model adapted to the peer-review process.
Submission, revision and acceptance of scientific publications are the main milestones
in this branching model, using git-tags and cross-links between data, software, and publications. This} ensures that the FAIR principles \citep{FAIR} are applied to a large extent in a challenging university research setting with an extremely low work overhead. 

The complete workflow further increases the sustainability of research software. Next to the known practice of issue tracking, our version of Test-Driven Development - adapted to CSE research software - gradually increases the test coverage of research software, allowing researchers to keep the focus on the subsequent scientific publication without introducing broad unit-testing unrelated to the research roadmap.
Automatic test quantification and result visualization using Jupyter notebooks in a data-processing pipeline provide a basis for discussions and data analysis that help in quickly identifying sources of errors.
Automatic testing combined with containerization and minimal workflow makes it possible to reproduce research results from any milestone across different computing platforms. 
Finally, the use of Continuous Integration for automating scientific workflows enables the results to be reproduced automatically from a git Web interface.
Granted, for researchers unfamiliar with the tools and techniques from the complete workflow, a non-negligible learning investment is expected.
However, this investment has a very high return rate. 
Increasing testing coverage using our version of TDD as well as automatic test quantification and visualization quickly pay off even for individual researchers and very small research groups. 
Continuous Integration increases research output more gradually once the time comes to combine existing research into new ideas.

\section*{Acknowledgements}
This work was funded by the Deutsche Forschungsgemeinschaft (DFG, German Research Foundation) -- Project-ID 265191195 -- SFB 1194, and by the Hessian LOEWE initiative within the Software-Factory 4.0 project.
The authors would also like to thank the Federal Government and the Heads of Government of the Länder, as well as the Joint Science
Conference (GWK), for their funding and support within the framework of the NFDI4Ing consortium. Funded by the German Research
Foundation (DFG) - project number 442146713.
Benchmarks were conducted on the Lichtenberg cluster at TU Darmstadt.

The authors are grateful to Moritz Schwarzmeier for his work on the NFDI4Ing Knowledgebase \citep{NFDI4IngKBase}, and to Moritz Schwarzmeier and Tobias Tolle for providing valuable constructive feedback on the RSE workflow and this manuscript.




\bibliographystyle{elsarticle-num-names} 
\bibliography{literature.bib}

\begin{thebibliography}{48}
\expandafter\ifx\csname natexlab\endcsname\relax\def\natexlab#1{#1}\fi
\providecommand{\url}[1]{\texttt{#1}}
\providecommand{\href}[2]{#2}
\providecommand{\path}[1]{#1}
\providecommand{\DOIprefix}{doi:}
\providecommand{\ArXivprefix}{arXiv:}
\providecommand{\URLprefix}{URL: }
\providecommand{\Pubmedprefix}{pmid:}
\providecommand{\doi}[1]{\href{http://dx.doi.org/#1}{\path{#1}}}
\providecommand{\Pubmed}[1]{\href{pmid:#1}{\path{#1}}}
\providecommand{\bibinfo}[2]{#2}
\ifx\xfnm\relax \def\xfnm[#1]{\unskip,\space#1}\fi
\bibitem[{Heroux et~al.(2020)Heroux, Gonsiorowski, Gupta, Milewicz, Moulton,
  Watson, Willenbring, Zamora, and Raybourn}]{2020:heroux:psip}
\bibinfo{author}{M.~A. Heroux}, \bibinfo{author}{E.~Gonsiorowski},
  \bibinfo{author}{R.~Gupta}, \bibinfo{author}{R.~Milewicz},
  \bibinfo{author}{J.~D. Moulton}, \bibinfo{author}{G.~R. Watson},
  \bibinfo{author}{J.~Willenbring}, \bibinfo{author}{R.~J. Zamora},
  \bibinfo{author}{E.~M. Raybourn},
\newblock \bibinfo{title}{Lightweight software process improvement using
  productivity and sustainability improvement planning (psip)},
\newblock in: \bibinfo{editor}{G.~Juckeland},
  \bibinfo{editor}{S.~Chandrasekaran} (Eds.), \bibinfo{booktitle}{Tools and
  Techniques for High Performance Computing}, \bibinfo{publisher}{Springer
  International Publishing}, \bibinfo{address}{Cham}, \bibinfo{year}{2020}, pp.
  \bibinfo{pages}{98--110}.
\bibitem[{FAI(2022)}]{FAIR}
\bibinfo{title}{{FAIR principles}},
  \bibinfo{howpublished}{\url{https://www.go-fair.org/fair-principles/}},
  \bibinfo{year}{2022}. \bibinfo{note}{Accessed: 2022-08-03}.
\bibitem[{Jiménez et~al.(2017)Jiménez, Kuzak, Alhamdoosh, Barker, Batut,
  Borg, Capella-Gutierrez, Chue~Hong, Cook, Corpas, Flannery, Garcia, Gelpí,
  Gladman, Goble, González~Ferreiro, Gonzalez-Beltran, Griffin, Grüning,
  Hagberg, Holub, Hooft, Ison, Katz, Leskošek, López~Gómez, Oliveira,
  Mellor, Mosbergen, Mulder, Perez-Riverol, Pergl, Pichler, Pope, Sanz,
  Schneider, Stodden, Suchecki, Svobodová~Vařeková, Talvik, Todorov,
  Treloar, Tyagi, van Gompel, Vaughan, Via, Wang, Watson-Haigh, and
  Crouch}]{Jimenez2017}
\bibinfo{author}{R.~C. Jiménez}, \bibinfo{author}{M.~Kuzak},
  \bibinfo{author}{M.~Alhamdoosh}, \bibinfo{author}{M.~Barker},
  \bibinfo{author}{B.~Batut}, \bibinfo{author}{M.~Borg},
  \bibinfo{author}{S.~Capella-Gutierrez}, \bibinfo{author}{N.~Chue~Hong},
  \bibinfo{author}{M.~Cook}, \bibinfo{author}{M.~Corpas},
  \bibinfo{author}{M.~Flannery}, \bibinfo{author}{L.~Garcia},
  \bibinfo{author}{J.~L. Gelpí}, \bibinfo{author}{S.~Gladman},
  \bibinfo{author}{C.~Goble}, \bibinfo{author}{M.~González~Ferreiro},
  \bibinfo{author}{A.~Gonzalez-Beltran}, \bibinfo{author}{P.~C. Griffin},
  \bibinfo{author}{B.~Grüning}, \bibinfo{author}{J.~Hagberg},
  \bibinfo{author}{P.~Holub}, \bibinfo{author}{R.~Hooft},
  \bibinfo{author}{J.~Ison}, \bibinfo{author}{D.~S. Katz},
  \bibinfo{author}{B.~Leskošek}, \bibinfo{author}{F.~López~Gómez},
  \bibinfo{author}{L.~J. Oliveira}, \bibinfo{author}{D.~Mellor},
  \bibinfo{author}{R.~Mosbergen}, \bibinfo{author}{N.~Mulder},
  \bibinfo{author}{Y.~Perez-Riverol}, \bibinfo{author}{R.~Pergl},
  \bibinfo{author}{H.~Pichler}, \bibinfo{author}{B.~Pope},
  \bibinfo{author}{F.~Sanz}, \bibinfo{author}{M.~V. Schneider},
  \bibinfo{author}{V.~Stodden}, \bibinfo{author}{R.~Suchecki},
  \bibinfo{author}{R.~Svobodová~Vařeková}, \bibinfo{author}{H.~A. Talvik},
  \bibinfo{author}{I.~Todorov}, \bibinfo{author}{A.~Treloar},
  \bibinfo{author}{S.~Tyagi}, \bibinfo{author}{M.~van Gompel},
  \bibinfo{author}{D.~Vaughan}, \bibinfo{author}{A.~Via},
  \bibinfo{author}{X.~Wang}, \bibinfo{author}{N.~S. Watson-Haigh},
  \bibinfo{author}{S.~Crouch},
\newblock \bibinfo{title}{Four simple recommendations to encourage best
  practices in research software},
\newblock \bibinfo{journal}{F1000Research} \bibinfo{volume}{6}
  (\bibinfo{year}{2017}) \bibinfo{pages}{1--15}.
  \DOIprefix\doi{10.12688/f1000research.11407.1}.
\bibitem[{Fehr et~al.(2016)Fehr, Heiland, Himpe, and Saak}]{Fehr2016}
\bibinfo{author}{J.~Fehr}, \bibinfo{author}{J.~Heiland},
  \bibinfo{author}{C.~Himpe}, \bibinfo{author}{J.~Saak},
\newblock \bibinfo{title}{Best practices for replicability, reproducibility and
  reusability of computer-based experiments exemplified by model reduction
  software},
\newblock \bibinfo{journal}{AIMS Mathematics} \bibinfo{volume}{1}
  (\bibinfo{year}{2016}) \bibinfo{pages}{261--281}.
  \DOIprefix\doi{10.3934/Math.2016.3.261}, \bibinfo{note}{arXiv: 1607.01191}.
\bibitem[{Wilson et~al.(2014)Wilson, Aruliah, Brown, Chue~Hong, Davis, Guy,
  Haddock, Huff, Mitchell, Plumbley, Waugh, White, and Wilson}]{Wilson2014}
\bibinfo{author}{G.~Wilson}, \bibinfo{author}{D.~A. Aruliah},
  \bibinfo{author}{C.~T. Brown}, \bibinfo{author}{N.~P. Chue~Hong},
  \bibinfo{author}{M.~Davis}, \bibinfo{author}{R.~T. Guy},
  \bibinfo{author}{S.~H. Haddock}, \bibinfo{author}{K.~D. Huff},
  \bibinfo{author}{I.~M. Mitchell}, \bibinfo{author}{M.~D. Plumbley},
  \bibinfo{author}{B.~Waugh}, \bibinfo{author}{E.~P. White},
  \bibinfo{author}{P.~Wilson},
\newblock \bibinfo{title}{Best {Practices} for {Scientific} {Computing}},
\newblock \bibinfo{journal}{PLoS Biology} \bibinfo{volume}{12}
  (\bibinfo{year}{2014}). \DOIprefix\doi{10.1371/journal.pbio.1001745},
  \bibinfo{note}{arXiv: 1210.0530}.
\bibitem[{Wilson et~al.(2017)Wilson, Bryan, Cranston, Kitzes, Nederbragt, and
  Teal}]{wilson_good_2017}
\bibinfo{author}{G.~Wilson}, \bibinfo{author}{J.~Bryan},
  \bibinfo{author}{K.~Cranston}, \bibinfo{author}{J.~Kitzes},
  \bibinfo{author}{L.~Nederbragt}, \bibinfo{author}{T.~K. Teal},
\newblock \bibinfo{title}{Good enough practices in scientific computing},
\newblock \bibinfo{journal}{PLOS Computational Biology} \bibinfo{volume}{13}
  (\bibinfo{year}{2017}) \bibinfo{pages}{e1005510}. \URLprefix
  \url{https://journals.plos.org/ploscompbiol/article?id=10.1371/journal.pcbi.1005510}.
  \DOIprefix\doi{10.1371/journal.pcbi.1005510}, \bibinfo{note}{publisher:
  Public Library of Science}.
\bibitem[{Stanisic et~al.(2015)Stanisic, Legrand, and
  Danjean}]{stanisic_effective_2015}
\bibinfo{author}{L.~Stanisic}, \bibinfo{author}{A.~Legrand},
  \bibinfo{author}{V.~Danjean},
\newblock \bibinfo{title}{An {Effective} {Git} {And} {Org}-{Mode} {Based}
  {Workflow} {For} {Reproducible} {Research}},
\newblock \bibinfo{journal}{ACM SIGOPS Operating Systems Review}
  \bibinfo{volume}{49} (\bibinfo{year}{2015}) \bibinfo{pages}{61--70}.
  \URLprefix \url{https://doi.org/10.1145/2723872.2723881}.
  \DOIprefix\doi{10.1145/2723872.2723881}.
\bibitem[{Anzt et~al.(2019)Anzt, Chen, Cojean, Dongarra, Flegar, Nayak,
  Quintana-Ortí, Tsai, and Wang}]{anzt_towards_2019}
\bibinfo{author}{H.~Anzt}, \bibinfo{author}{Y.-C. Chen},
  \bibinfo{author}{T.~Cojean}, \bibinfo{author}{J.~Dongarra},
  \bibinfo{author}{G.~Flegar}, \bibinfo{author}{P.~Nayak},
  \bibinfo{author}{E.~S. Quintana-Ortí}, \bibinfo{author}{Y.~M. Tsai},
  \bibinfo{author}{W.~Wang},
\newblock \bibinfo{title}{Towards {Continuous} {Benchmarking}: {An} {Automated}
  {Performance} {Evaluation} {Framework} for {High} {Performance} {Software}},
\newblock in: \bibinfo{booktitle}{Proceedings of the {Platform} for {Advanced}
  {Scientific} {Computing} {Conference}}, {PASC} '19,
  \bibinfo{publisher}{Association for Computing Machinery},
  \bibinfo{address}{New York, NY, USA}, \bibinfo{year}{2019}, pp.
  \bibinfo{pages}{1--11}. \URLprefix
  \url{https://doi.org/10.1145/3324989.3325719}.
  \DOIprefix\doi{10.1145/3324989.3325719}.
\bibitem[{Riesch et~al.(2020)Riesch, Nguyen, and
  Jirauschek}]{riesch_bertha_2020}
\bibinfo{author}{M.~Riesch}, \bibinfo{author}{T.~D. Nguyen},
  \bibinfo{author}{C.~Jirauschek},
\newblock \bibinfo{title}{bertha: {Project} skeleton for scientific software},
\newblock \bibinfo{journal}{PLOS ONE} \bibinfo{volume}{15}
  (\bibinfo{year}{2020}) \bibinfo{pages}{e0230557}. \URLprefix
  \url{https://journals.plos.org/plosone/article?id=10.1371/journal.pone.0230557}.
  \DOIprefix\doi{10.1371/journal.pone.0230557}, \bibinfo{note}{publisher:
  Public Library of Science}.
\bibitem[{Sampedro et~al.(2018)Sampedro, Holt, and
  Hauser}]{sampedro_continuous_2018}
\bibinfo{author}{Z.~Sampedro}, \bibinfo{author}{A.~Holt},
  \bibinfo{author}{T.~Hauser},
\newblock \bibinfo{title}{Continuous {Integration} and {Delivery} for {HPC}:
  {Using} {Singularity} and {Jenkins}},
\newblock in: \bibinfo{booktitle}{Proceedings of the {Practice} and
  {Experience} on {Advanced} {Research} {Computing}}, {PEARC} '18,
  \bibinfo{publisher}{Association for Computing Machinery},
  \bibinfo{address}{New York, NY, USA}, \bibinfo{year}{2018}, pp.
  \bibinfo{pages}{1--6}. \URLprefix
  \url{https://doi.org/10.1145/3219104.3219147}.
  \DOIprefix\doi{10.1145/3219104.3219147}.
\bibitem[{BSS(2022)}]{BSSw}
\bibinfo{title}{{Better Scientific Software (BSSw)}},
  \bibinfo{howpublished}{\url{https://bssw.io/}}, \bibinfo{year}{2022}.
  \bibinfo{note}{Accessed: 2022-08-03}.
\bibitem[{NFD(2022{\natexlab{a}})}]{NFDI}
\bibinfo{title}{{National Research Data Initiative}},
  \bibinfo{howpublished}{\url{https://nfdi4ing.de/}},
  \bibinfo{year}{2022}{\natexlab{a}}. \bibinfo{note}{Accessed: 2022-08-03}.
\bibitem[{NFD(2022{\natexlab{b}})}]{NFDI4IngKBase}
\bibinfo{title}{{NFDI4Ing Knowledge Base}},
  \bibinfo{howpublished}{https://nfdi4ing.pages.rwth-aachen.de/knowledge-base/},
  \bibinfo{year}{2022}{\natexlab{b}}. \bibinfo{note}{Accessed: 2022-08-03}.
\bibitem[{SSI(2022)}]{SSI}
\bibinfo{title}{{Software Sustainability Institute}},
  \bibinfo{howpublished}{https://software.ac.uk/}, \bibinfo{year}{2022}.
  \bibinfo{note}{Accessed: 2022-08-03}.
\bibitem[{Community(2022)}]{TuringWay}
\bibinfo{author}{T.~T.~W. Community}, \bibinfo{title}{{The Turing Way: A
  handbook for reproducible, ethical and collaborative research}},
  \bibinfo{year}{2022}. \URLprefix
  \url{https://doi.org/10.5281/zenodo.6909298}.
  \DOIprefix\doi{10.5281/zenodo.6909298}.
\bibitem[{Sin(2022)}]{Singularity}
\bibinfo{title}{{Singularity}},
  \bibinfo{howpublished}{\url{https://singularity.lbl.gov/}},
  \bibinfo{year}{2022}. \bibinfo{note}{Accessed: 2022-08-03}.
\bibitem[{Khuvis et~al.(2019)Khuvis, You, Na, Brozell, Franz, Dockendorf,
  Gardiner, and Tomko}]{Khuvis2019}
\bibinfo{author}{S.~Khuvis}, \bibinfo{author}{Z.~Q. You},
  \bibinfo{author}{H.~Na}, \bibinfo{author}{S.~Brozell},
  \bibinfo{author}{E.~Franz}, \bibinfo{author}{T.~Dockendorf},
  \bibinfo{author}{J.~Gardiner}, \bibinfo{author}{K.~Tomko},
\newblock \bibinfo{title}{A continuous integration-based framework for software
  management},
\newblock in: \bibinfo{booktitle}{{ACM} {International} {Conference}
  {Proceeding} {Series}}, \bibinfo{year}{2019}.
  \DOIprefix\doi{10.1145/3332186.3332219}.
\bibitem[{Jup(2022)}]{Jupyter}
\bibinfo{title}{{Jupyter}}, \bibinfo{howpublished}{\url{https://jupyter.org/}},
  \bibinfo{year}{2022}. \bibinfo{note}{Accessed: 2022-08-03}.
\bibitem[{Beg et~al.(2021)Beg, Taka, Kluyver, Konovalov, Ragan-Kelley, Thiery,
  and Fangohr}]{Beg2021}
\bibinfo{author}{M.~Beg}, \bibinfo{author}{J.~Taka},
  \bibinfo{author}{T.~Kluyver}, \bibinfo{author}{A.~Konovalov},
  \bibinfo{author}{M.~Ragan-Kelley}, \bibinfo{author}{N.~M. Thiery},
  \bibinfo{author}{H.~Fangohr},
\newblock \bibinfo{title}{Using {Jupyter} for {Reproducible} {Scientific}
  {Workflows}},
\newblock \bibinfo{journal}{Computing in Science and Engineering}
  \bibinfo{volume}{23} (\bibinfo{year}{2021}) \bibinfo{pages}{36--46}.
  \DOIprefix\doi{10.1109/MCSE.2021.3052101}.
\bibitem[{Bin(????)}]{Binder}
\bibinfo{title}{{Binder}}, \bibinfo{howpublished}{https://mybinder.org/}, ????
  \bibinfo{note}{Accessed: 2022-05-05}.
\bibitem[{Chacon and Straub(2014)}]{Chacon2014}
\bibinfo{author}{S.~Chacon}, \bibinfo{author}{B.~Straub}, \bibinfo{title}{Pro
  git}, \bibinfo{publisher}{Apress}, \bibinfo{year}{2014}.
\bibitem[{Git(2022{\natexlab{a}})}]{GitLab}
\bibinfo{title}{{GitLab}}, \bibinfo{howpublished}{\url{https://gitlab.com}},
  \bibinfo{year}{2022}{\natexlab{a}}. \bibinfo{note}{Accessed: 2022-08-03}.
\bibitem[{Git(2022{\natexlab{b}})}]{GitHub}
\bibinfo{title}{{GitHub}}, \bibinfo{howpublished}{\url{https://github.com}},
  \bibinfo{year}{2022}{\natexlab{b}}. \bibinfo{note}{Accessed: 2022-08-03}.
\bibitem[{Bit(2022)}]{Bitbucket}
\bibinfo{title}{{Bitbucket}},
  \bibinfo{howpublished}{\url{https://bitbucket.org}}, \bibinfo{year}{2022}.
  \bibinfo{note}{Accessed: 2022-08-03}.
\bibitem[{Git(2022{\natexlab{a}})}]{Gitflow}
\bibinfo{title}{{Gitflow}},
  \bibinfo{howpublished}{\url{https://nvie.com/posts/a-successful-git-branching-model/}},
  \bibinfo{year}{2022}{\natexlab{a}}. \bibinfo{note}{Accessed: 2022-08-03}.
\bibitem[{Git(2022{\natexlab{b}})}]{Githubflow}
\bibinfo{title}{{Github flow}},
  \bibinfo{howpublished}{\url{https://guides.github.com/introduction/flow/}},
  \bibinfo{year}{2022}{\natexlab{b}}. \bibinfo{note}{Accessed: 2022-08-03}.
\bibitem[{Zen(2022)}]{Zenodo}
\bibinfo{title}{{Zenodo}}, \bibinfo{howpublished}{\url{https://zenodo.org/}},
  \bibinfo{year}{2022}. \bibinfo{note}{Accessed: 2022-08-03}.
\bibitem[{TUd(2022)}]{TUdatalib}
\bibinfo{title}{{TUdatalib}},
  \bibinfo{howpublished}{\url{https://tudatalib.ulb.tu-darmstadt.de/}},
  \bibinfo{year}{2022}. \bibinfo{note}{Accessed: 2022-08-03}.
\bibitem[{ArX(2022)}]{ArXiv}
\bibinfo{title}{{arXiv}}, \bibinfo{howpublished}{\url{https://arxiv.org/}},
  \bibinfo{year}{2022}. \bibinfo{note}{Accessed: 2022-08-03}.
\bibitem[{Sugimori et~al.(1977)Sugimori, Kusunoki, Cho, and
  UCHIKAWA}]{Sugimori1977}
\bibinfo{author}{Y.~Sugimori}, \bibinfo{author}{K.~Kusunoki},
  \bibinfo{author}{F.~Cho}, \bibinfo{author}{S.~UCHIKAWA},
\newblock \bibinfo{title}{Toyota production system and kanban system
  materialization of just-in-time and respect-for-human system},
\newblock \bibinfo{journal}{The international journal of production research}
  \bibinfo{volume}{15} (\bibinfo{year}{1977}) \bibinfo{pages}{553--564}.
\bibitem[{Beck(2003)}]{Beck2003}
\bibinfo{author}{K.~Beck}, \bibinfo{title}{Test-driven development: by
  example}, \bibinfo{publisher}{Addison-Wesley Professional},
  \bibinfo{year}{2003}.
\bibitem[{{The HDF Group}(2010)}]{HDF5}
\bibinfo{author}{{The HDF Group}}, \bibinfo{title}{{Hierarchical data format
  version 5}}, \bibinfo{year}{2000-2010}. \URLprefix
  \url{http://www.hdfgroup.org/HDF5}.
\bibitem[{Kurtzer et~al.(2017)Kurtzer, Sochat, and Bauer}]{Kurtzer2017}
\bibinfo{author}{G.~M. Kurtzer}, \bibinfo{author}{V.~Sochat},
  \bibinfo{author}{M.~W. Bauer},
\newblock \bibinfo{title}{Singularity: Scientific containers for mobility of
  compute},
\newblock \bibinfo{journal}{PloS one} \bibinfo{volume}{12}
  (\bibinfo{year}{2017}) \bibinfo{pages}{e0177459}.
\bibitem[{Fowler and Foemmel(2006)}]{Fowler2006}
\bibinfo{author}{M.~Fowler}, \bibinfo{author}{M.~Foemmel},
  \bibinfo{title}{Continuous integration}, \bibinfo{year}{2006}.
\bibitem[{MWE(2022)}]{MWErepo}
\bibinfo{title}{{Minimal Working Example Repository}}, \bibinfo{year}{2022}.
  \URLprefix
  \url{https://gitlab.com/cse-ci-examples/cse-rse-workflow-code/-/tree/2022-07-21-preprint},
  \bibinfo{note}{accessed: 2022-07-21}.
\bibitem[{Maric(2022{\natexlab{a}})}]{ResearchReport}
\bibinfo{author}{T.~Maric}, \bibinfo{title}{Computational science and
  engineering - research software engineering workflow : Research report},
  \bibinfo{year}{2022}{\natexlab{a}}. \URLprefix
  \url{https://doi.org/10.48328/tudatalib-921.2}.
\bibitem[{Maric(2022{\natexlab{b}})}]{Data}
\bibinfo{author}{T.~Maric}, \bibinfo{title}{Computational science and
  engineering - research software engineering workflow : Research data},
  \bibinfo{year}{2022}{\natexlab{b}}. \URLprefix
  \url{https://doi.org/10.48328/tudatalib-910.2}.
\bibitem[{Maric(2022{\natexlab{c}})}]{CodeSnapshot}
\bibinfo{author}{T.~Maric}, \bibinfo{title}{Computational science and
  engineering - research software engineering workflow : Research code},
  \bibinfo{year}{2022}{\natexlab{c}}. \URLprefix
  \url{https://tudatalib.ulb.tu-darmstadt.de/handle/tudatalib/3524.5}.
\bibitem[{geo(2022)}]{geophase}
\bibinfo{title}{{Git repository of the geophase library}},
  \bibinfo{howpublished}{https://git.rwth-aachen.de/leia/geophase},
  \bibinfo{year}{2022}. \bibinfo{note}{Accessed: 2022-08-03}.
\bibitem[{Mari\'{c}(2020)}]{Maric2020ccs}
\bibinfo{author}{T.~Mari\'{c}}, \bibinfo{title}{Iterative volume-of-fluid
  interface positioning in general polyhedrons with consecutive cubic spline
  interpolation}, \bibinfo{year}{2020}.
  \href{http://arxiv.org/abs/2012.02227}{{\tt arXiv:2012.02227}}.
\bibitem[{Mari{\'c} et~al.(2018)Mari{\'c}, Marschall, and Bothe}]{Maric2018}
\bibinfo{author}{T.~Mari{\'c}}, \bibinfo{author}{H.~Marschall},
  \bibinfo{author}{D.~Bothe},
\newblock \bibinfo{title}{An enhanced un-split face-vertex flux-based vof
  method},
\newblock \bibinfo{journal}{Journal of computational physics}
  \bibinfo{volume}{371} (\bibinfo{year}{2018}) \bibinfo{pages}{967--993}.
\bibitem[{Mari{\'c} et~al.(2020)Mari{\'c}, Kothe, and Bothe}]{Maric2020rev}
\bibinfo{author}{T.~Mari{\'c}}, \bibinfo{author}{D.~B. Kothe},
  \bibinfo{author}{D.~Bothe},
\newblock \bibinfo{title}{Unstructured un-split geometrical volume-of-fluid
  methods--a review},
\newblock \bibinfo{journal}{Journal of Computational Physics}
  \bibinfo{volume}{420} (\bibinfo{year}{2020}) \bibinfo{pages}{109695}.
\bibitem[{TUG(2022)}]{TUGitLab}
\bibinfo{title}{{TUGitLab}},
  \bibinfo{howpublished}{\url{https://git.rwth-aachen.de/}},
  \bibinfo{year}{2022}. \bibinfo{note}{Accessed: 2022-08-03}.
\bibitem[{CMa(2022)}]{CMake}
\bibinfo{title}{{CMake}}, \bibinfo{howpublished}{\url{https://cmake.org}},
  \bibinfo{year}{2022}. \bibinfo{note}{Accessed: 2022-08-03}.
\bibitem[{geo(2022)}]{geophaseCI}
\bibinfo{title}{{CI results repository of the geophase library}},
  \bibinfo{howpublished}{https://leia.pages.rwth-aachen.de/geophase-ci-results/},
  \bibinfo{year}{2022}. \bibinfo{note}{Accessed: 2022-08-03}.
\bibitem[{Lehr et~al.(2018)Lehr, H\"{u}ck, and Bischof}]{2018:lehr:pira}
\bibinfo{author}{J.-P. Lehr}, \bibinfo{author}{A.~H\"{u}ck},
  \bibinfo{author}{C.~Bischof},
\newblock \bibinfo{title}{Pira: Performance instrumentation refinement
  automation},
\newblock in: \bibinfo{booktitle}{5th ACM SIGPLAN Intl. Workshop on Artificial
  Intelligence and Empirical Methods for Software Engineering and Parallel
  Computing Systems}, AI-SEPS 2018, \bibinfo{publisher}{ACM},
  \bibinfo{year}{2018}, pp. \bibinfo{pages}{1--10}. \URLprefix
  \url{http://doi.acm.org/10.1145/3281070.3281071}.
  \DOIprefix\doi{10.1145/3281070.3281071}.
\bibitem[{Lehr et~al.(2019)Lehr, Calotoiu, Bischof, and
  Wolf}]{2019:lehr:piraII}
\bibinfo{author}{J.-P. Lehr}, \bibinfo{author}{A.~Calotoiu},
  \bibinfo{author}{C.~Bischof}, \bibinfo{author}{F.~Wolf},
\newblock \bibinfo{title}{Automatic instrumentation refinement for empirical
  performance modeling},
\newblock in: \bibinfo{booktitle}{2019 IEEE/ACM Intl. Workshop on Programming
  and Performance Visualization Tools (ProTools)},
  \bibinfo{organization}{IEEE}, \bibinfo{year}{2019}, pp.
  \bibinfo{pages}{40--47}. \DOIprefix\doi{10.1109/ProTools49597.2019.00011}.
\bibitem[{{Sokolowski} et~al.(2020){Sokolowski}, Lehr, Bischof, and
  Salvaneschi}]{2020:sokolowski:hycloud}
\bibinfo{author}{D.~{Sokolowski}}, \bibinfo{author}{J.-P. Lehr},
  \bibinfo{author}{C.~Bischof}, \bibinfo{author}{G.~Salvaneschi},
\newblock \bibinfo{title}{Leveraging hybrid cloud hpc with multitier reactive
  programming},
\newblock in: \bibinfo{booktitle}{2020 IEEE/ACM International Workshop on
  Interoperability of Supercomputing and Cloud Technologies (SuperCompCloud)},
  \bibinfo{year}{2020}. \bibinfo{note}{To appear}.

\end{thebibliography}


\end{document}